\documentclass[sn-mathphys-num]{sn-jnl}

\usepackage{graphicx}%
\usepackage{multirow}%
\usepackage{amsmath,amssymb,amsfonts}%
\usepackage{amsthm}%
\usepackage{mathrsfs}%
\usepackage[title]{appendix}%
\usepackage{xcolor}%
\usepackage{textcomp}%
\usepackage{manyfoot}%
\usepackage{booktabs}%
\usepackage{algorithm}%
\usepackage{algorithmicx}%
\usepackage{algpseudocode}%
\usepackage{listings}%

\theoremstyle{thmstyleone}%
\theoremstyle{thmstyletwo}%

\theoremstyle{thmstylethree}%

\begin{document}

\title[Article Title]{Near Infrared Quantum Ghost Spectroscopy for Threats Detection}


\author*[1]{\fnm{Andrea} \sur{Chiuri}}\email{andrea.chiuri@enea.it}

\author[1]{\fnm{Federico} \sur{Angelini}}\email{federico.angelini@enea.it}

\author[3]{\fnm{Ilaria} \sur{Gianani}}\email{ilaria.gianani@uniroma3.it}

\author[1]{\fnm{Simone} \sur{Santoro}}\email{simone.santoro@enea.it}

\author[2,3]{\fnm{Marco} \sur{Barbieri}}\email{marco.barbieri@uniroma3.it}

\affil[1]{\orgdiv{Nuclear Department}, \orgname{ENEA}, \orgaddress{\street{Via E. Fermi 45}, \city{Frascati}, \postcode{00044}, \country{Italy}}}

\affil[2]{\orgdiv{Dipartimento di Scienze}, \orgname{Universit\`{a} degli Studi Roma Tre}, \orgaddress{\street{Via della Vasca Navale 84}, \city{Rome}, \postcode{00146}, \country{Italy}}}

\affil[3]{\orgdiv{Istituto Nazionale di Ottica}, \orgname{CNR}, \orgaddress{\street{Largo E. Fermi 6}, \city{Florence}, \postcode{50125}, \country{Italy}}}


\abstract{Quantum Sensing is a rapidly growing branch of research within the area of quantum science and technology offering key resources, beyond classical ones, with potential for commercialisation of novel (quantum) sensors. The exploitation of quantum resources offered by photons can boost the performance of quantum sensors for innovative and challenging applications. In this paper we build on the idea of the Quantum Ghost Spectroscopy (QGS), i.e. the counterpart in the frequency domain of Quantum Ghost Imaging (QGI), targeting specific applications in the detection of possible threats. This is implemented by exploiting the opportunities offered by Quantum Optics, i.e. the generation of photon pairs characterized by spectral correlations. We will discuss our main results obtained with pure QGS experiments showing that it is possible to assess the presence of a target dealing with a low resources measurement. The time-frequency domain reveals a huge potential for several applications and frequency correlations represent a versatile tool that can be exploited to enable the spectral analysis of objects where a direct measurement would not be feasible (e.g. security). The use of nondegenerate sources of correlated photons allowed to reveal spectral features in the near infrared wavelengths employing the usual detectors for the visible region.}




\maketitle

\section{Introduction}\label{Sec:Intro}

Detection of threats is a key security issue that encompasses diverse scenarios, from logistics to on-field operations, and should rely on highly efficient and robust sensors capable of operating remotely. Indeed, adopting appropriate diagnostic tools can be a necessary course of action in order to design strategies to protect operators, civilians and infrastructures, based on a data-informed evaluation of the risks. 
In this respect, improving the capabilities of reliable and accurate sensors for chemical, biological, radiological, and nuclear (CBRN) threats such as harmful gases and volatile organic compounds (VOCs), nuclear wastes and post-explosion residues, allows to provide a more carefully planned security strategy. However, implementing measurements in safe conditions - by either remote sensing or operation by automated mobile units - contrasts with the need of incorporating high-performance equipment for the detection suitable for species identification. This is further aggravated by the necessity of accessing information in specific domains, such as the infrared, where detection usually has high costs and low-performance. This is a fundamental roadblock whether samples are either sensed remotely or collected on the scene and analysed in the laboratory. 
Correlation measurements offer a promising path to overcome these obstacles: by exploiting spectral correlations, one can perform measurements in a convenient spectral range while accessing information on the desired yet technologically more demanding one~\cite{Kalashnikov16}.  A key technique which enables both correlation measurements and remote sensing is that of the so-called ``ghost" measurements~\cite{padgett17ptrsa}. 

Ghost imaging (GI) is a sophisticated method to image an object without spatially-resolving the light it interacts with, and can be realised either by exploiting classical \cite{Valencia05,Janassek18,bennink04prl} or quantum correlations \cite{Pittman95,padgett17ptrsa,Aspden13,Aspden15,Morris15}. The framework of Quantum Ghost Imaging (QGI) makes use spatial quantum correlations of pairs of photons typically generated via Spontaneous Parametric Down Conversion (SPDC). SPDC-generated photons can however exhibit correlations as well in other degrees of freedoms, such as frequency and polarization, thus allowing to extend the same paradigm to those domains. Specifically when looking at the frequency domain, if the SPDC source generates pairs of photons in a nondegenerate configuration, so that the energy of the incoming photon is not divided equally among the generated ones, it is be possible trough Quantum Ghost Spectroscopy (QGS) to link infrared (IR) and visible components of the emission in such a way that spectral information in the former range can be assessed by looking only at the latter. This means that the requirements for remote-yet-accurate measurements are shifted from the IR to the visible domain, for which much more solid, reliable, and cost-effective solutions are available. One of the major benefits of QGS compared to its classical counterpart is that it provides better signal-to-noise ratio (SNR) \cite{PhysRevX.4.011049,sullivan10pra}, however due to the long acquisition times it is particularly helpful when the full spectrum or a complete lineshape is not necessary but a spectral feature could be enough to arise an alert about the presence of a possible harmful substance. This makes the technique an optimal match for CBRN detection \cite{chierici20}, while due to higher SNR allows to work with overtones and absorption in the  near IR range ($\lambda \simeq$ 900-1600 nm), where the peaks are not strong as in the mid IR ($\lambda\geq 2\, \mu$m) . The latter spectral region has been explored in a quantum scenario to exploit the capabilities offered by this novel approach to overcome the technical drawbacks affecting the classical techniques, e.g. the Fourier-transform infrared spectroscopy (FTIR). The proposed approaches in a quantum scenario rely mainly on non-degenerate SPDC sources \cite{Arahata21,Cai24,Kalashnikov16,Kalachev_2008} and nonlinear quantum interferometers \cite{Paterova20,Neves24,Tashima:24, Cardoso:24, Gili22, Grafe22}, i.e. measurements with undetected photons. They certainly represent an interesting perspective for the quantum technologies but this spectral region is affected by several drawbacks, e.g. the stability and reliability of the interferometer, the atmosphere shows a strong absorbance.  

Telecom wavelengths, around. 1550nm,  represent a useful spectral range and a helpful solution for  some of these issues with a realistic perspective through the remote sensing implementation. Interestingly, there are features in this region allowing to identify different species of contaminants with a single device. Our work is devoted to benchmarking the operation of such an equipment based on available technology. Our goal is then to observe what realistic observations may be carried out over the spectrum of the SPDC which is typically broader of individual absorption features. A reliable signal is obtained nevertheless, and can provide guidelines for adoption in threat detection. We thus aim at supporting a realistic technology transfer to pave the way to a new generation of optical quantum sensors and metrology devices based on quantum enhanced spectroscopy and imaging. 


\section{Quantum Ghost Spectroscopy}\label{Sec:QGS}

\subsection{State of the art}

The idea of exploiting correlations in SPDC photon pairs for imaging purposes was first demonstrated in ~\cite{Pittman95}, using raster scanning. This has been refined through the adoption of  advanced single-photon cameras allowing detection even with a limited number of runs \cite{Aspden13,Aspden15,Morris15}, hence paving the way to the application of QGI in microscopy \cite{Aspden15,Morris15} and sub-shot noise imaging \cite{Samantaray2017}. The presence of quantum correlations also allows to consider robustness in the presence of an eavesdropper compromising  the object's image \cite{Malik12}.  Gatti and co-workers \cite{Valencia05} demonstrated that pseudo-thermal light can be used in place of  down-conversion, introducing a new line of practical imaging techniques. Several results have appeared concerning the use of (pseudo-) thermal sources and the comparison between classical and quantum approaches for non-degenerate and degenerate wavelength configurations \cite{Meda_2017,Chan09}. A different approach considers computational GI \cite{PhysRevA.78.061802}: it is performed by using a spatial light modulator creating random intensity patterns with which to illuminate the object. Spatial correlations are now established between an optical beam and a calculable pattern held in the computer memory.  In this case, only a single-element detector is required to measure the intensity of the light that is transmitted or backscattered by the object. The final image can be simply obtained by summing all of the patterns, each weighted by the signal from the detector.

If the bucket detector of conventional GI is replaced by a detector resolving the spatial spectrum of the signal photons, the 3D geometry of an object can be obtained~\cite{Dangelo16}. This scheme has been experimentally demonstrated with a classical light source \cite{Pepe17} and holds promise for the realization of low-illumination power 3D imaging. This 3D GI has been also studied and realized implementing a setup based on asynchronous single photon timing using Single Photon Avalanche Detectors (SPADs). This scheme enables photon pairing with arbitrary path length difference and does, therefore, obviate the dependence on optical delay lines \cite{Walter19,Pitsch:21,Pitsch23}

\begin{figure*}[h!]
    \centering
    \includegraphics[width=0.6 \textwidth]{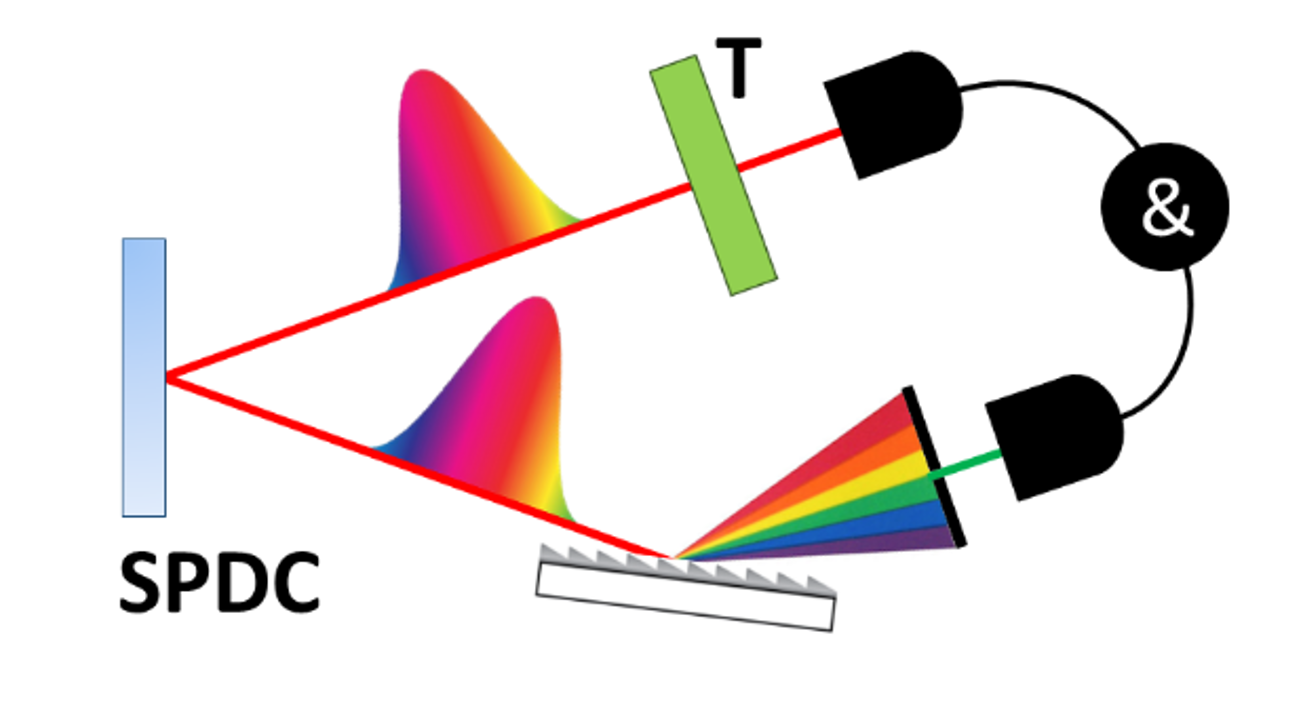}
    \caption{Quantum Ghost Spectroscopy (QGS).  Correlated emission from SPDC is used both to illuminate the target object $T$ and to reach a frequency-resolved detector. By means of the correlation measurement, by means {\it e.g.} of coincidence detection, the modification induced by $T$ on the spectrum is inferred even if there is no energy resolution available at the target location. This models scenarios where it would be difficult to transport complex equipment, while pieces with modest SWaP can still be moved.}
    \label{fig:ghost}
\end{figure*}

The adoption of ghost techniques for spectroscopy have been introduced in \cite{Scarcelli,Yabushita04}, with demonstrations based on  frequency-entangled photons from SPDC, as shown in Fig.~\ref{fig:ghost}. The absorption spectrum of a suitable sample was measured by counting coincidences. In \cite{Janassek18} the authors reported experimental implementations of classical GS using different sources of thermal light based on photon bunching. The use has then been extended to resource-efficient recognition of absorption features~\cite{PhysRevA.105.013506} and to the joint use of spatial and spectral degrees of freedom\cite{doi:10.1021/acsphotonics.3c01108}. Also, in \cite{Cai24} the authors tested this kind of approach with commercial fibers (10 km). In the same paper they worked with a nondegenerate source (NIR-MIR) and the photon emitted in the MIR region was upconverted to the NIR range.

The ``ghost" approach, either for classical or quantum light, can be extended to different configurations. There have been demonstration with orbital angular momentum \cite{Leach10}, and polarisation \cite{Chirkin18,Magnitskiy20}, extending to the imaging of birefringent targets. Temporal ghost imaging has been used, e.g. to transmit images through single-mode fibers \cite{Dong2016}. 
Ghost imaging  has also been implemented with atoms in the classical \cite{Khakimov2016}  as well as in the quantum regime~\cite{Chan:09}; notably, this latter work has demonstrated that high-order correlation functions ($n > 2$) can yield to improvements in image reconstruction. 

\subsection{Benefits from Quantum Ghost Spectroscopy}

Employing quantum correlation could allow to perform remote sensing measurements, akin to ghost imaging protocols \cite{Walter19,Pitsch:21,Pitsch23}. This can be particularly advantageous when the objects at hand are not easily accessible due to the suspected presence of  CBRN threats. In these instances, it is vital to extract information at a distance, both to ensure the safety of the users and for ease of measurement operations. Although this is also possible using classical spectral correlations, using quantum ones can show better performance \cite{PhysRevA.105.013506,sullivan10pra,bennink04prl,padgett17ptrsa}.

Foremost, the use of quantum correlated photons increases the SNR and enables measurements that are more robust against external noise (e.g. due stray light, scattering).  An analysis of the SNR in the quantum case $S_Q$ and in the traditional case $S_T$
gives  $S_Q/S_T =\eta_i/\left(N_i\Delta t\right)$, where $\eta_i$ is the detection efficiency for the idler channel, $N_i$ is the number of noise photon counts in the idler arm, and $\Delta t$ is the coincidence time window of the coincidence circuit~\cite{PhysRevX.4.011049}. Therefore, QGS brings about an SNR comparable to classical techniques while using lower intensities. The number of needed photons may be further reduced by introducing techniques such as compressed sensing.  Hence, this strategy can show better performance with respect to classical ones when low-intensity signals are in use.  

Case in point, QGS represents an optimal solution when dealing with systems that strongly affect the transmission resulting in low signal rates.  By exploiting the quantum correlations, the detection can be made less sensitive to turbulence and visibility of speckles than traditional one~\cite{Walter19}. Previous work \cite{PhysRevA.109.042617} demonstrated a viable route for fast spectral discrimination: the more different the spectral profile after the absorption, the lesser resources are needed. Classical sensors require more resources (e.g. time, measurements) and high-flux of radiation are required with the risk to damage the sample.  In this regime, quantum strategies could achieve a better uncertainty when the imaging problem is cast as  parameter estimation~\cite{PhysRevA.105.013506}. QGS may thus allow spectroscopy with minimal illumination, implying it could guarantee an eye-safe deployment. The latter represents a mandatory requirement whenever real scenarios are dealt with~\cite{Walter19}. As this approach is based on random emission and random direction,  it could guarantee the desired non-detectability as long as the positioning of the bucket detector permits.

SPDC sources can generate pairs of photons belonging to different wavelength ranges (the first in the visible and the second in the IR range), making it possible to link different spectral regimes. This means  the requirements for accurate measurements could be shifted from the IR to the visible domain, for which much more solid, reliable and cost-effective solutions are available. In addition, the non-degenerate approach could allow to combine superior atmospheric transmission and reduced scattering of IR spectral range.

The two photons could be used for independent measurements, although only one interacts with the sample. This means that two different types of information can be obtained simultaneously, thus maximizing the gained amount of information from the available photons. Incorporating other degrees of freedom (spatial domain or the polarization) would be helpful in determining not only the presence of a threat but also its position and size. This could be achieved by merging QGS and QGI obtaining high resolution in both the spatial and the spectral/spatio-spectral domain \cite{doi:10.1021/acsphotonics.3c01108} without sacrificing photons by filtering or enlarging the needed bandwidth. In this case “only” two detectors are sufficient. Notice that in classical multispectral imaging, a narrow spectral filter is employed for each wavelength channel absorbing the other spectral components or many images are taken simultaneously, thus necessitating many detectors.

\section{Methods}\label{Sec:Exp}

Our test system is based on the QGS apparatus extensively described in the Appendix and schematised in Fig.~\ref{fig:schememain}. Photon pairs were produced by illuminating a nonlinear cristal (NLC): the idler photon at lower energy ($\lambda_{s}\simeq1550$ nm) went through the sample and then reached the bucket detector, a SPAD. The signal photon ($\lambda_{i}\simeq 810$ nm) is delivered to a single-grating spectrometer followed by an intensified CCD (iCCD). The spectral acquisition is triggered by the SPAD, and, in order to account for the delay in the bucket detector response, the signal photon is delayed by means of an optical delay line in a multimode fibre, while fine tuning is implemented by means of an FPGA. The thickness of the NLC was chosen to be either 1 mm or 3 mm depending on the sample hence on its absorption features, since this dictates the spectral width of the generated photons, while the correlation mainly depends on the pump shape. The grating in the spectrometer was selected accordingly to have 600 lines/mm or 1200 lines/mm by means of an automated control.
\begin{figure*}[h!]
    \centering
    \includegraphics[width=\textwidth]{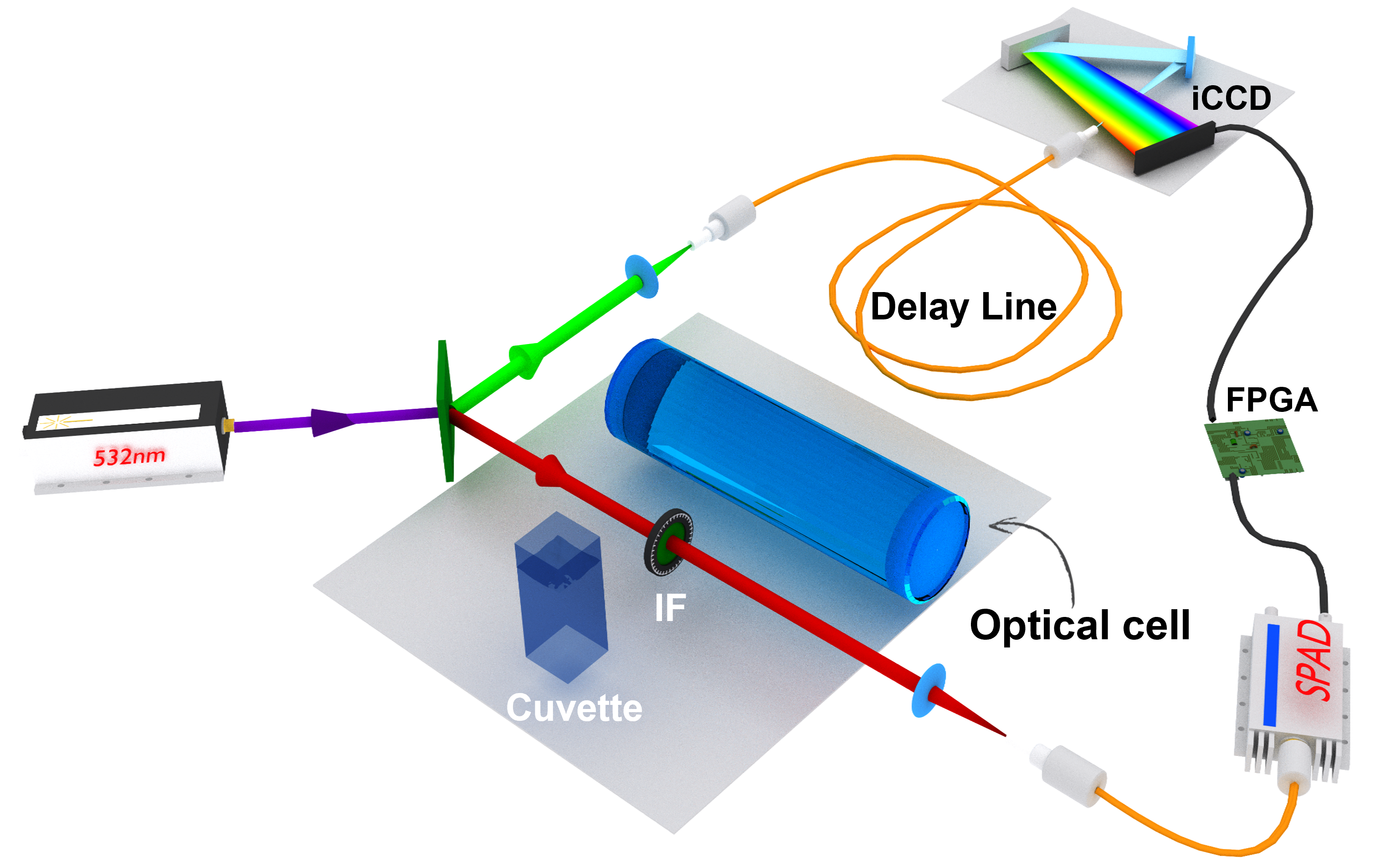}
    \caption{Schematics of the experiment. All the possible configurations of the idler arm have been included, i.e. cuvette (optical cell) for liquid (gaseous) samples and the IF filter to calibrate the device.}
    \label{fig:schememain}
\end{figure*}

The QGS setup has been calibrated by using an interference filter, and then employed with both liquid and gaseous samples. For the liquid samples a cuvette was placed instead of the test filter and the same setup has been employed. For gaseous samples, we designed and realized an optical cell (1 m length, 50.8 mm width) equipped with a suitable vacuum system composed by a membrane pump for the prevacuum and a turbomolecular pump. This cell was integrated in the bucket arm to allow the interaction between the sample and the idler photon (See Appendix). 

We performed experiments based on the absorbance of the samples and the intensity of the collected radiation can be described by a simple Lambert-Beer equation:
\begin{equation}
I=I_0 e^{- \mu (\lambda) \Delta l}
\end{equation}
where $I$ $\left (I_0\right)$ is the intensity with (without) the sample, $\mu(\lambda)$ is the absorption coefficient and $\Delta l$ is the length of the absorbing material.  Transmittance and absorbance, defined as $T=I / I_0$ and $A=-\ln (T)$, respectively, can be obtained directly from the measured spectral profiles with and without the sample (blank).

\section{Results}

\subsection{Volatile Organic Compounds}
VOCs are organic chemicals that evaporate into the atmosphere at room temperature.  This category embrace a large number of compounds: hydrocarbon VOCs having both hydrogen and carbon atoms that include benzene and toluene; oxygenate VOCs containing carbon, hydrogen, and oxygen, that are the result of car exhausts and atmospheric chemical reactions. VOCs comprise many substances for daily use, e.g. gasoline, paints and lacquers, cleaning supplies, pesticides, building materials, and glues and adhesives; also the manufacturing or processing of these substances may emit VOCs. Large manufacturing plants using solvents or toxic chemicals during production can release dangerous levels of VOCs in the air. Common toxic VOCs, include: acetone, acetylene, benzene, cyclohexene, ethanol, formaldehyde, methanol, and propanol. We focused on liquid Ethanol and Dichloromethane (DCM), as they bear distinctive features in the NIR region and were compatible with the security regulamentations of our institutions.


\begin{figure*}
    \centering
    \includegraphics[width=\textwidth]{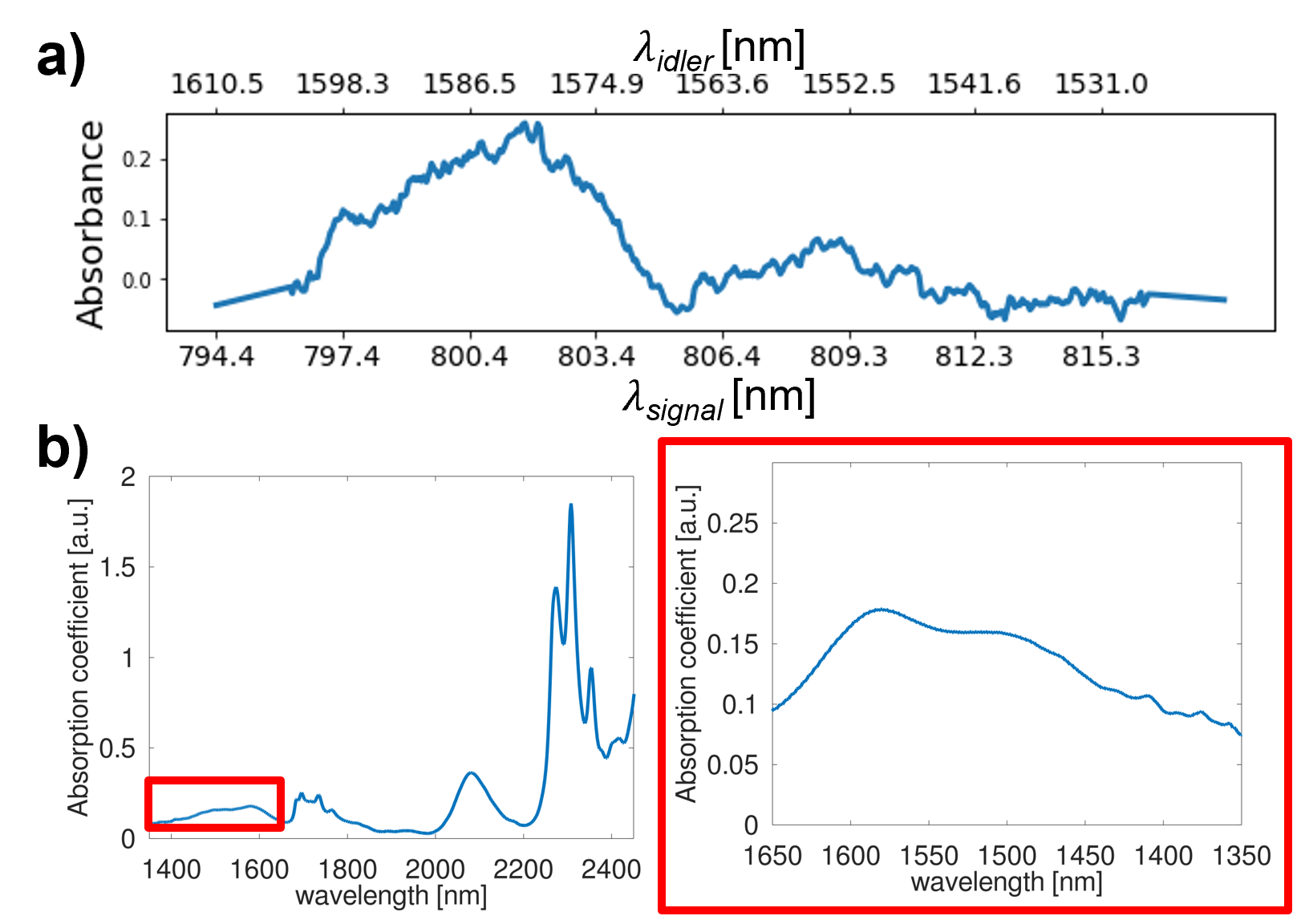}
    \caption{a) Measured absorbance showing a behaviour close to the expected one, as reported in b). Notice that the wavelength axis refers to the signal photon, hence it is actually reversed for the idler. b) Reference near IR spectrum of liquid ethanol as obtained by the National Institute of Standards and Technology (NIST) database and confirmed by those reported in literature \cite{Chowdhury21}. The imaginary part of the refractive index, available on this database, can be considered as the absorption coefficient.}
    \label{fig:ethanol}
\end{figure*}

\textit{Ethanol - } We have used a sample of pure ethanol contained in a cuvette with $\Delta l =0.3$ cm thickness due to the high absorbance of the substance. \cite{Chowdhury21,CZECHLOWSKI201914}.  We have adopted a 1 mm thickness of NLC in order to obtain wider bandwidth from the source tuned to the broad bands characterising the spectrum. This was then analysed with the 600 lines/mm grating. Figure~\ref{fig:ethanol}
reports the measured absorbance, presenting a broad peak which can be attributed to $OH$ bounds by comparison with the literature.

\begin{figure*}
    \centering
    \includegraphics[width=\textwidth]{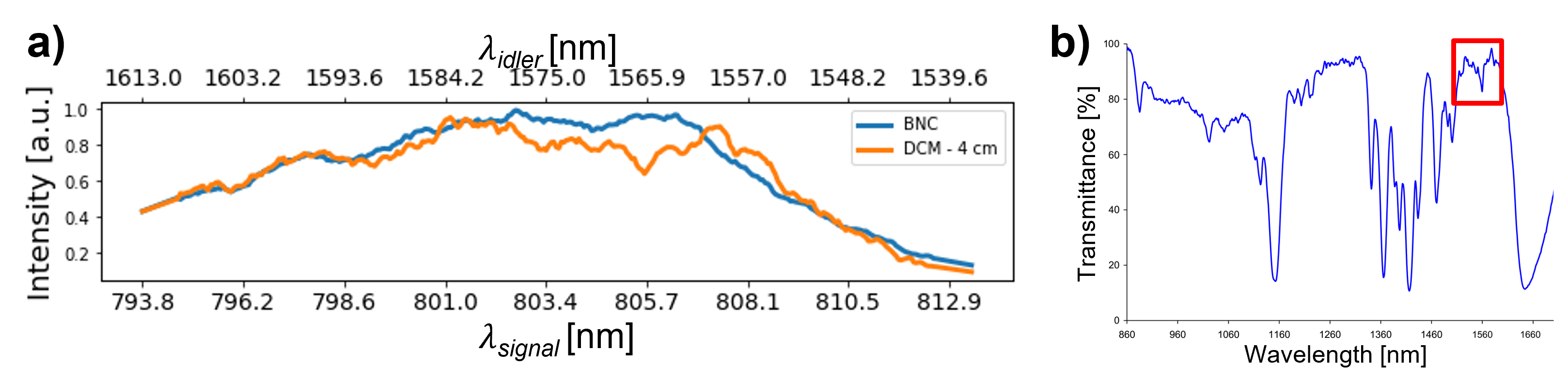}
    \caption{Dichloromethane near IR spectrum. a) \textit{Orange}: measurement performed filling a cuvette with few ml of pure DCM. \textit{Blue}: Reference spectrum measured with an empty cuvette. b) Reference as available online on open-source repositories and in literature \cite{DCM}. This spectrum has been acquired with spectrophotometer composed by a Halogen light bulb, i.e. the light source, and a Ocean Optics near IR (NIR-512) temperature-regulated InGaAs detector spectrometer with IR fiber optic light guide. The liquid sample has been allocated in a tiny beaker, i.e. $\approx 20ml$ of DCM into $\approx 2cm$ liquid optical path. The blank has been subtracted and the transmittance normalized to 100$\%$. }
    \label{fig:dcm}
\end{figure*}

\textit{Dichloromethane (DCM) - } The same experimental conditions have been employed for  DCM in Figure~\ref{fig:schemeapp}),  but in this case $\Delta l = 4$ cm because of the lower cross section of this sample. In this case, the NIR absorption spectrum shows complex overlapping overtones of MIR absorption features and we have investigated the region in the red box  reported in Figure~\ref{fig:dcm}. The measurement in the quantum ghost configuration roughly corresponds to the theoretical spectrum. 

\subsection{Explosives}
A further class of possible threats is represented by  explosives, and the use of such dangerous substances is not limited to military action scenarios. Several molecules show interesting bands in the NIR region, e.g. acetylene, acetone peroxide (TATP), dynamite (nitro-glycerine), ammonium nitrate, TNT , etc. We focused on acetylene as the representative of these compounds, because it is often employed in thefts and robberies. 

\textit{Acetylene - } This sample is in the gaseous form, hence we have employed the optical cell kept at a pressure close to the atmospheric conditions. The length of the NLC is 3 mm in order to increase the brightness of the source, and the spectrum has been analysed with the 1200 lines/mm grating. Such gaseous samples are characterized by very narrow peaks with specific line-shapes depending on several external parameters, e.g. pressure, temperature \cite{Swann00,nist2001}. All these spectral features in the NIR region of the Acetylene, i.e. rotational–vibrational bands, cannot be detected due to the a limited spectral resolution but out results allow nevertheless to identify a specific absorbance also for this target.  

\begin{figure*}[h!]
    \centering
    \includegraphics[width=\textwidth]{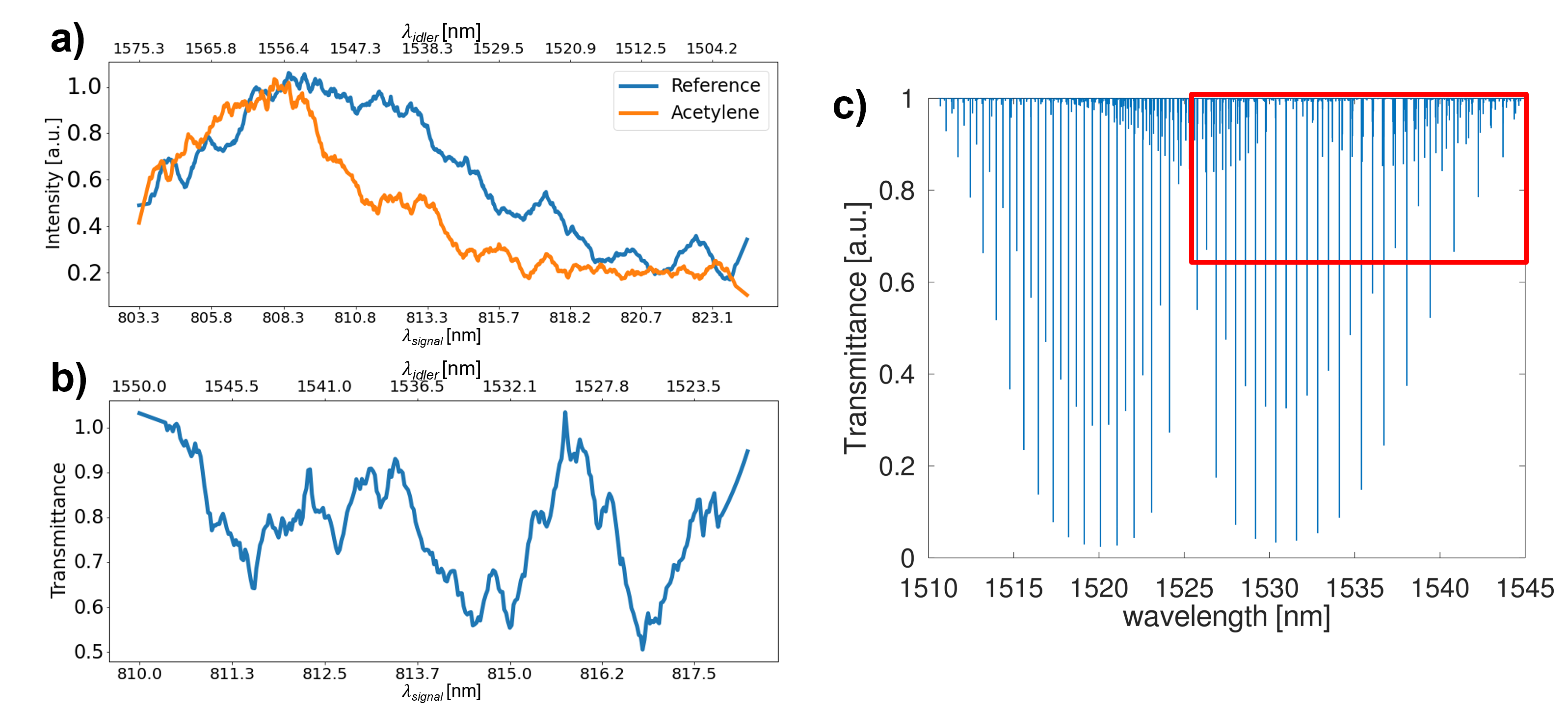}
    \caption{Acetylene near IR spectrum. 
    a) Orange/Blue: Measurement with/without the gaseous sample and a double-passage in the optical cell at $\approx 1160$ hPa. b) Corresponding Transmittance. c) NIR spectrum obtained with a simulation on the HITRAN database \cite{hitran} ($T: 300 K$, $P: 0.001 atm$, optical path: 2 m) which is consistent with those reported in literature \cite{Swann00,nist2001}. The telecom wavelengths under investigation in the red box show several narrow peaks to be considered. These can be discriminated at low pressure and broader peaks should be obtained with a different configuration, i.e. higher pressure and temperature \cite{Swann00,nist2001}. A convolution of these broad peaks should characterize the spectrum acquired with the realized quantum ghost spectrometer.}
    \label{fig:acetylene}
\end{figure*}

\subsection{Carbon Dioxide}
Carbon Dioxide ($CO_2$) has been considered because of its implications for many aspects related to the security. Along with methane $CH_4$, it is one of the bulk gases produced by waste degradation \cite{Shen81}.  The detection of $CO_2$ and $CH_4$ could be related to the presence of nuclear waste. Indeed, radiocarbon (C-14) is a long-lived radionucleide that represents one of the main source of radioactive gas emissions in nuclear facilities, mostly in the form of carbon dioxide or methane, produced during the biodegradation of radioactive waste \cite{Fleisher2017}. It is a harmful gas for the human health and it is involved in many issues related to the clime and the environment.

\begin{figure*}[h!]
    \centering
    \includegraphics[width=\textwidth]{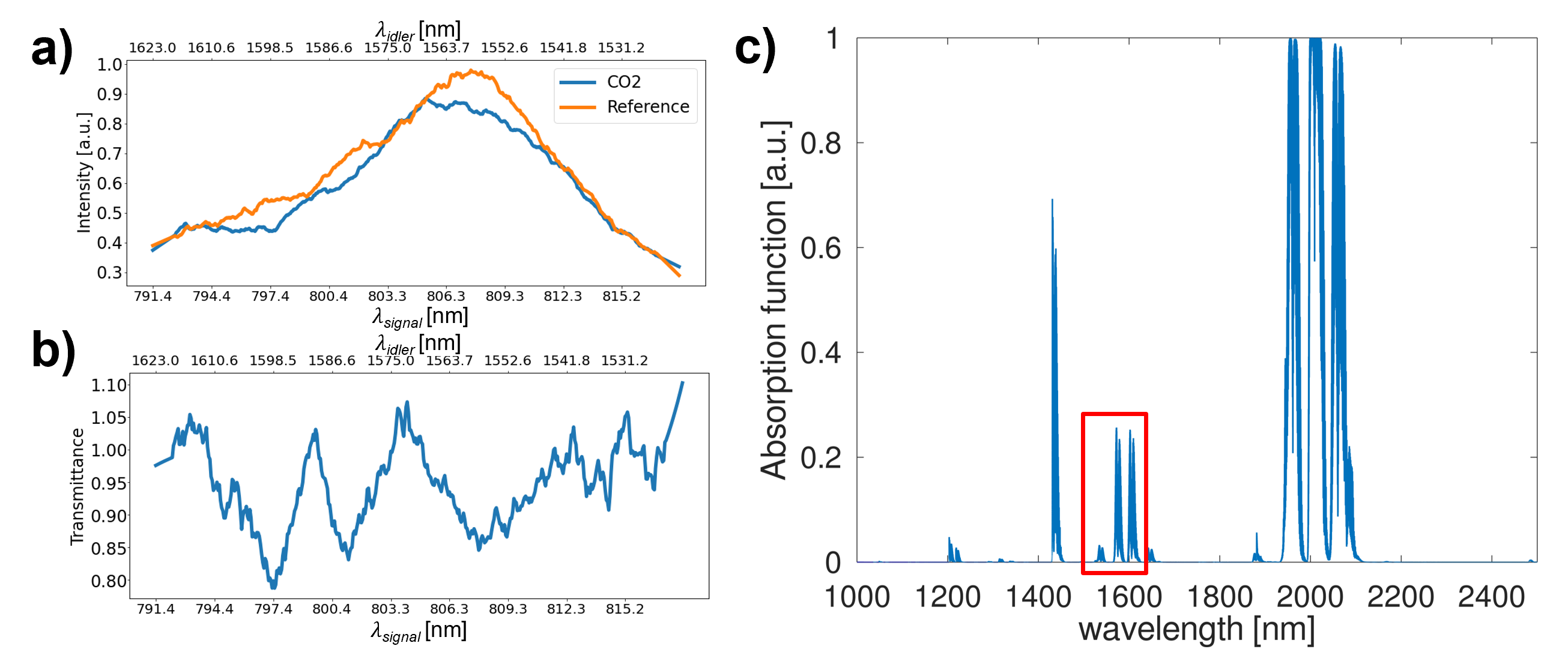}
    \caption{$CO_2$ near IR spectrum. a) Blue/Orange: Measurement with/without the gaseous sample and a double-passage in the optical cell kept at $\approx 2200$ hPa.  b) Corresponding transmittance obtained from th ghost spectroscopy data. c) NIR spectrum obtained with a simulation on the HITRAN database \cite{hitran} ($T: 300 K$, $P: 2 atm$, optical path: 2 m) which is consistent with those reported in literature \cite{rs13224502}. The telecom wavelengths under investigation in the red box show several peaks to be considered.}
    \label{fig:co2}
\end{figure*}

The experiment has been conducted as reported in Figure~\ref{fig:schemeapp}c and a double passage in the optical cell was necessary to obtain the spectrum reported in Figure~\ref{fig:co2}. For this target the 1 mm  NLC was necessary to cover the whole spectral region of interest. Our measurements  reveal features which do ressembled those observed on reference spectra.

\section{Discussion}\label{Sec:Disc}

The spectra of all the substances we have investigated are able to reveal the presence of the contaminants. However, while the features can be related to  those observed in reference measurements, the matching is far from perfect: this can be attributed to the limited resolution of the spectrometer, which, for instance, can not resolve individual lines, as well as to the measurement conditions themselves. We have shown what the spectra are like when employing a spectrometer whose resolution is similar to current portable devices,  except for the different requirements in single-photon detection. This highlights the importance of accurately characterise the response of the device to various physical parameters affecting the measurement, notably temperature and pressure~\cite{A903398J,Swann00,nist2001}. Reference spectra can be useful for the interpretation of the resulting spectra, but have limited applicability to tasks such as automated recognition. If such systems are sought after, then their training should be based on the actual experimental response.

The main disadvantage of the technique is in its low count rates and long image acquisition times. These are mainly limited by the technologies of the frequency resolving single photon detector in the signal arm and the bucket detectors in the idler arm. Nevertheless this field sees rapid technological improvements with an impressive growth in the last decade, e.g.  SPAD arrays \cite{Bruschini2019,Madonini21} , TimePix cameras \cite{Fisher-Levine_2016, NOMEROTSKI201926, Ianzano2020, Vidyapin2023}, and Superconducting Nanowires Single Photon Detector arrays \cite{Resta2023,Hao2024,Wollman:19}, although these are not as appealing for transportable solutions.

\section{Conclusion}\label{Sec:Conc}
We have illustrated characteristic examples of  the employ of QGS for the detection of substances associated to threats, either in liquid or gaseous specimens.  We demonstrated that, despite the low signal rates, their presence can be ascertained by observing distinctive spectral features, although these do not match exactly those indicated in reference spectra. 

QGS could be an optimal solution when dealing with systems that strongly affect the transmission resulting in exceedingly low signal rates, which are those that would otherwise require more effort to reconstruct their full lineshape. The CBRN threats could be included in this kind of targets and in realistic scenarios they do not require the full identification of all the peaks composing their spectra. Some spectral features could be enough to gather the necessary information and launch an alert. The QGS approach offers also the possibility to work with low radiation levels and this is relevant when the target could be an unknown harmful substance. The discussed technique has showed also a great spectral tunability as it can take advantage of the non-degenerate measurements, enabling the investigation in otherwise hardly accessible spectral regimes.

\backmatter

\bmhead{Acknowledgements}
The authors acknowledge Isabella Giardina and Gaetano Terranova for preparing the samples.

\bmhead{Funding}
This work was supported by the NATO Science for Peace and Security (SPS) Programme, project HADES (id G5839).

\bmhead{Author contribution}
A.C., M.B., F.A. and S.S. performed the measurements, M.B. and I.G. were involved in planning and supervised the
work, A.C. processed the experimental data, performed the analysis, drafted the manuscript and
designed the figures. F.A., S.S., M.B., I.G. aided in interpreting
the results and worked on the manuscript. All authors discussed the results and commented on the
manuscript. 

\bmhead{Data availability}
The datasets generated during and/or analysed during the current study are available from the corresponding author on reasonable request.

\begin{appendices}

\section{The Quantum Ghost Spectrometer}\label{Appendix}

\begin{figure*}[h!]
    \centering
    \includegraphics[width=\textwidth]{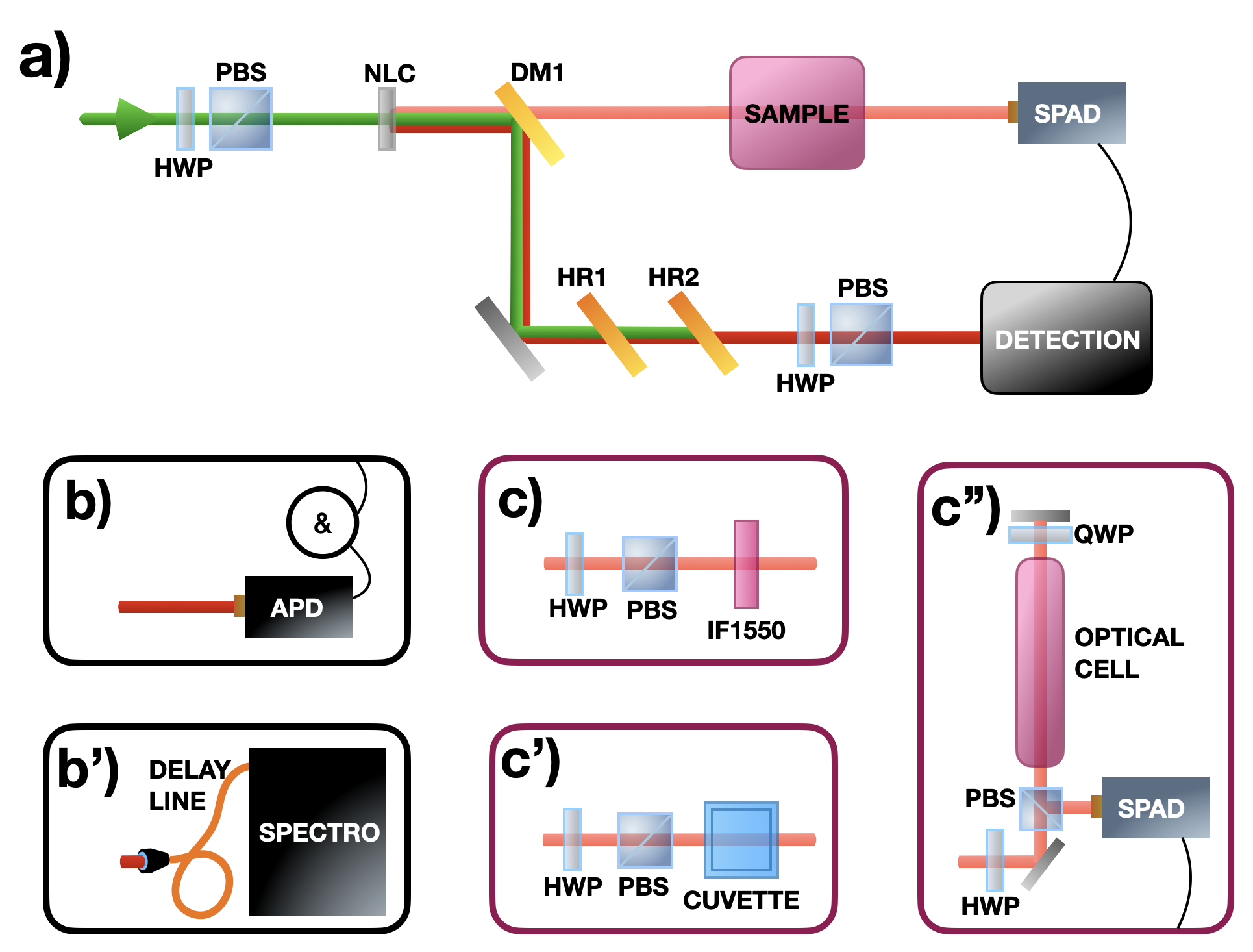}
    \caption{Experimental setup used to perform the measurements discussed in the main text. The detection box (black) has been arranged to optimize the coincidences (b) or to perform frequency-resolved measurements (b'). The sample box (purple) has been suitably adapted to the specific application: c) calibration with a bandpass filter, c') liquid samples, c'') gaseous samples. Here the double passage in the optical cell has been implemented including a quarter waveplate rotated at $45^{\circ}$ and a mirror. The reflected photons are polarized V and reflected by the PBS towards the SPAD. NLC: nonlinear crystal, DM: dichoric mirror, PBS: polarizing beam splitter, IF: bandpass interference filters, HR: high-reflectance mirrors at 532 nm, $\lambda/2$ ($\lambda/4$): half(quarter) waveplate.}
    \label{fig:schemeapp}
\end{figure*}

Our source adopts a pulsed laser at $\lambda_p$=532 nm as the pump beam. Its parameters are: average power up to 200 mW, repetition rate 40 MHz, pulse width 8 ps. This guaranteed excellent frequency correlation, as the pump spectrum is much narrower than the resolution of our spectrometer (cfr. infra). The NLC was a $\beta$-barium borate bulk slab, generating signal-idler photon pairs by type-I phase matching at $\lambda_{s} \approx 810nm$ and $\lambda_{i} \approx 1550nm$, respectively, in the collinear regime. This is shown in Fig.~\ref{fig:schemeapp}a.  Both photons are collected by means of spatially multimode fibres.

By suitably tilting the NLC and and employing crystals characterized by different thickness, {\it i.e. 1 mm and 3 mm}, the source could be suitably adapted to explore broad spectral ranges, e.g. 1450-1650 nm which is interesting for detecting several analytes. Indeed,  for a type-I NLC, the phase matching conditions write $n(\lambda_p,\theta ')/\lambda_p=n_o (\lambda_s)/\lambda_s+n_o (\lambda_i)/\lambda_i$ , where $n_o (\lambda) (n_e (\lambda))$ is the refractive index of ordinary (extraordinary) light, and $\theta '$ is the angle between the propagation direction of light and the optical axis in the crystal. The choice of the dispersion of the material and the orientation of its crystal axes with respect to the laser direction offers a wide range of tunability for the emitted photons in terms of wavelength, longitudinal momentum, and polarization.

Concerning the detection, the idler was sent to the bucket detector, i.e. a SPAD (MPD PDM-IR), after a dichroic mirror (DM) which reflects the residual pump and the signal. This was triggered by the sync-out TTL emitted by the laser to lower the Dark Count Rate (DCR) of the bucket arm. The signal is filtered by means of two high-reflectance mirrors, HR1 and HR2, in order to reflect the residual pump, and allow operation without bandpass filters before the detectors. For alignment purposes, the signal photon was coupled to an avalanche photodiode (APD),  Fig.~\ref{fig:schemeapp}b., and then the same fibre was branched into the spectrometer,  Fig.~\ref{fig:schemeapp}b'. 

The frequency resolved measurement was realized by a spectrometer composed by a spectrograph (Andor Kymera 328i) equipped with two gratings (600 or 1200 lines/mm) and an iCCD (Andor iStar DH334T-18U-73). The bucket detector (i.e. SPAD) was connected to a Field Programmable Gate Array (FPGA) board suitably programmed to compress the output TTL and introduce a variable delay. For each exposure window, the intensifier of the iCCD camera on the signal arm was opened by the TTL coming from the FPGA as a Direct Gate allowing to reduce the noise thanks to a narrow gate width. The variable delay was necessary to optimize the synchronization between signal and idler photon detection with a simple and elegant solution. This fine adjustment was necessary to compensate the electronic delays and the delay line introduced on the signal arm, i.e. 15 m of multi-mode fiber corresponding to $\approx 60$ ns. 

\begin{figure*}[h!]
    \centering
    \includegraphics[width=\textwidth]{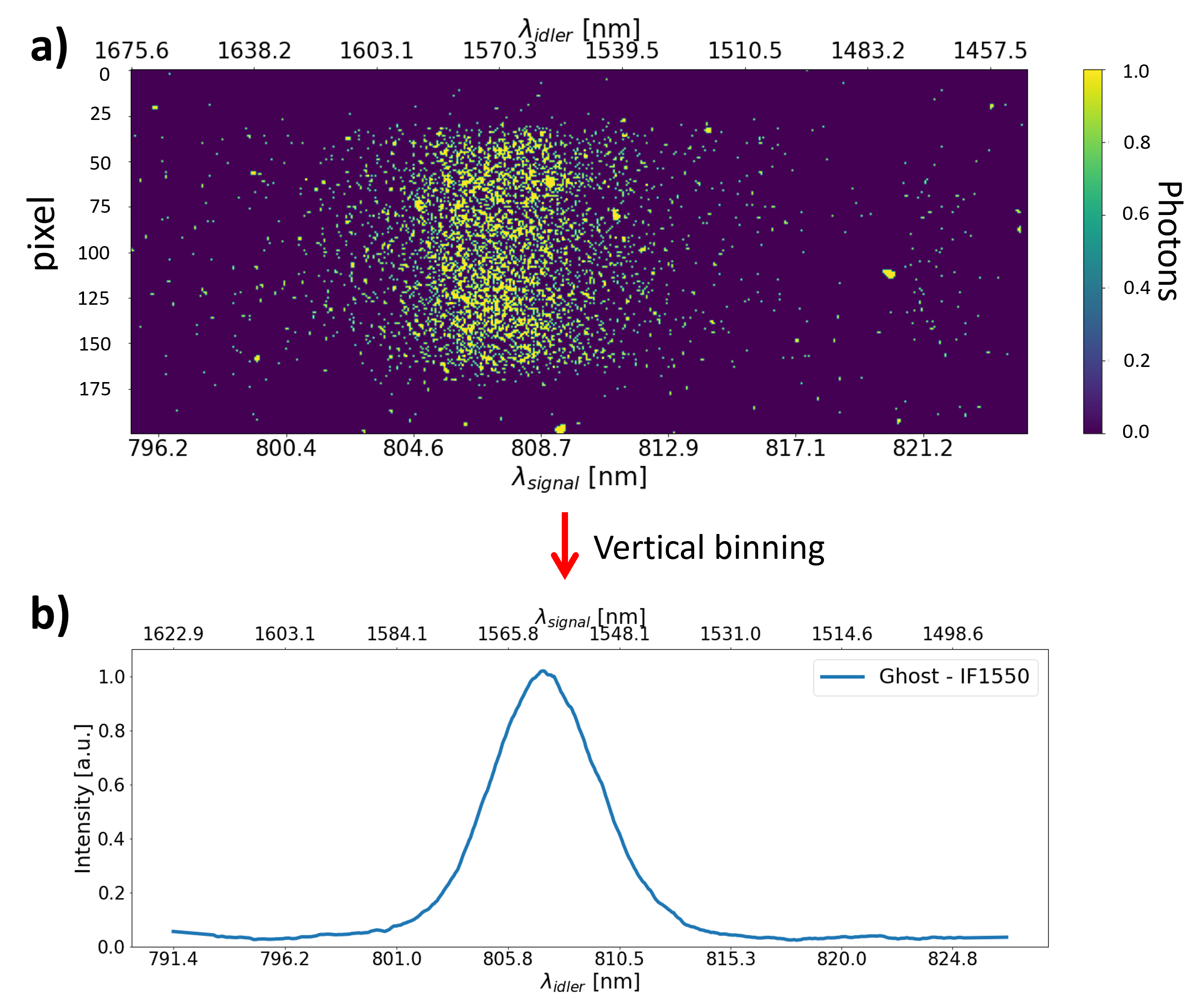}
    \caption{Frequency-resolved measurements acquired in t = 1800 s (Exposure time=9s, Accumulation=200 frames). a) raw counts. b) integrating over the spatial shape, i.e.vertical binning, the spectrum characterizing the specific target can be obtained.}
    \label{fig:appendix2}
\end{figure*}

The iCCD outputs a spatially resolved image over the vertical dimension y, while the orthogonal direction x is devoted to the spectral analysis. The raw outcomes of our measurements are thus the number of counts from the iCCD for every pixel acquired in the photon-counting regime - uncertainties can then be associated to the Poisson statistics of the retrieved counts. The spectral discrimination ability of our setup can be assessed by introducing objects with different spectral responses. In our work, we investigated several targets with different transmission/absorption profiles. For each measurement the integration time was $\approx 3600 s$ but this value is strongly affected by the low brightness of the source based on BBO crystals. The latter can be certainly improved by employing  different choices, e.g. \cite{Tashima:24,Arahata21,Kalashnikov16}. A usual result of a measurement is presented in Figure~\ref{fig:appendix2}, in which we show the raw data obtained when inserting the calibration interference filter (IF) on the idler beam. The final spectra have been obtained by integrating over the $y$ direction and employing a smoothing algorithm, {\it i.e.} a Savitzky–Golay filter.

\end{appendices}


\bibliography{sn-bibliography}


\begin{thebibliography}{61}
\ifx \bisbn   \undefined \def \bisbn  #1{ISBN #1}\fi
\ifx \binits  \undefined \def \binits#1{#1}\fi
\ifx \bauthor  \undefined \def \bauthor#1{#1}\fi
\ifx \batitle  \undefined \def \batitle#1{#1}\fi
\ifx \bjtitle  \undefined \def \bjtitle#1{#1}\fi
\ifx \bvolume  \undefined \def \bvolume#1{\textbf{#1}}\fi
\ifx \byear  \undefined \def \byear#1{#1}\fi
\ifx \bissue  \undefined \def \bissue#1{#1}\fi
\ifx \bfpage  \undefined \def \bfpage#1{#1}\fi
\ifx \blpage  \undefined \def \blpage #1{#1}\fi
\ifx \burl  \undefined \def \burl#1{\textsf{#1}}\fi
\ifx \doiurl  \undefined \def \doiurl#1{\url{https://doi.org/#1}}\fi
\ifx \betal  \undefined \def \betal{\textit{et al.}}\fi
\ifx \binstitute  \undefined \def \binstitute#1{#1}\fi
\ifx \binstitutionaled  \undefined \def \binstitutionaled#1{#1}\fi
\ifx \bctitle  \undefined \def \bctitle#1{#1}\fi
\ifx \beditor  \undefined \def \beditor#1{#1}\fi
\ifx \bpublisher  \undefined \def \bpublisher#1{#1}\fi
\ifx \bbtitle  \undefined \def \bbtitle#1{#1}\fi
\ifx \bedition  \undefined \def \bedition#1{#1}\fi
\ifx \bseriesno  \undefined \def \bseriesno#1{#1}\fi
\ifx \blocation  \undefined \def \blocation#1{#1}\fi
\ifx \bsertitle  \undefined \def \bsertitle#1{#1}\fi
\ifx \bsnm \undefined \def \bsnm#1{#1}\fi
\ifx \bsuffix \undefined \def \bsuffix#1{#1}\fi
\ifx \bparticle \undefined \def \bparticle#1{#1}\fi
\ifx \barticle \undefined \def \barticle#1{#1}\fi
\bibcommenthead
\ifx \bconfdate \undefined \def \bconfdate #1{#1}\fi
\ifx \botherref \undefined \def \botherref #1{#1}\fi
\ifx \url \undefined \def \url#1{\textsf{#1}}\fi
\ifx \bchapter \undefined \def \bchapter#1{#1}\fi
\ifx \bbook \undefined \def \bbook#1{#1}\fi
\ifx \bcomment \undefined \def \bcomment#1{#1}\fi
\ifx \oauthor \undefined \def \oauthor#1{#1}\fi
\ifx \citeauthoryear \undefined \def \citeauthoryear#1{#1}\fi
\ifx \endbibitem  \undefined \def \endbibitem {}\fi
\ifx \bconflocation  \undefined \def \bconflocation#1{#1}\fi
\ifx \arxivurl  \undefined \def \arxivurl#1{\textsf{#1}}\fi
\csname PreBibitemsHook\endcsname

\bibitem[\protect\citeauthoryear{Kalashnikov et~al.}{2016}]{Kalashnikov16}
\begin{barticle}
\bauthor{\bsnm{Kalashnikov}, \binits{D.}},
\bauthor{\bsnm{Paterova}, \binits{A.}},
\bauthor{\bsnm{Kulik}, \binits{S.}},
\bauthor{\bsnm{Krivitsky}, \binits{A.} \bsuffix{Leonid}}:
\batitle{Infrared spectroscopy with visible light}.
\bjtitle{Nature Photonics}
\bvolume{10},
\bfpage{98}--\blpage{101}
(\byear{2016})
\doiurl{10.1038/nphoton.2015.2529}
\end{barticle}
\endbibitem

\bibitem[\protect\citeauthoryear{Padgett and Boyd}{2017}]{padgett17ptrsa}
\begin{barticle}
\bauthor{\bsnm{Padgett}, \binits{M.J.}},
\bauthor{\bsnm{Boyd}, \binits{R.W.}}:
\batitle{An introduction to ghost imaging: quantum and classical}.
\bjtitle{Philosophical Transactions of the Royal Society A: Mathematical, Physical and Engineering Sciences}
\bvolume{375}(\bissue{2099}),
\bfpage{20160233}
(\byear{2017})
\doiurl{10.1098/rsta.2016.0233}
{\href{https://arxiv.org/abs/https://royalsocietypublishing.org/doi/pdf/10.1098/rsta.2016.0233}{{https://royalsocietypublishing.org/doi/pdf/10.1098/rsta.2016.0233}}}
\end{barticle}
\endbibitem

\bibitem[\protect\citeauthoryear{Valencia et~al.}{2005}]{Valencia05}
\begin{barticle}
\bauthor{\bsnm{Valencia}, \binits{A.}},
\bauthor{\bsnm{Scarcelli}, \binits{G.}},
\bauthor{\bsnm{D'Angelo}, \binits{M.}},
\bauthor{\bsnm{Shih}, \binits{Y.}}:
\batitle{Two-photon imaging with thermal light}.
\bjtitle{Phys. Rev. Lett.}
\bvolume{94},
\bfpage{063601}
(\byear{2005})
\doiurl{10.1103/PhysRevLett.94.063601}
\end{barticle}
\endbibitem

\bibitem[\protect\citeauthoryear{Janassek et~al.}{2018}]{Janassek18}
\begin{botherref}
\oauthor{\bsnm{Janassek}, \binits{P.}},
\oauthor{\bsnm{Herdt}, \binits{A.}},
\oauthor{\bsnm{Blumenstein}, \binits{S.}},
\oauthor{\bsnm{Elsäßer}, \binits{W.}}:
Ghost spectroscopy with classical correlated amplified spontaneous emission photons emitted by an erbium-doped fiber amplifier.
Applied Sciences
\textbf{8}(10)
(2018)
\doiurl{10.3390/app8101896}
\end{botherref}
\endbibitem

\bibitem[\protect\citeauthoryear{Bennink et~al.}{2004}]{bennink04prl}
\begin{barticle}
\bauthor{\bsnm{Bennink}, \binits{R.S.}},
\bauthor{\bsnm{Bentley}, \binits{S.J.}},
\bauthor{\bsnm{Boyd}, \binits{R.W.}},
\bauthor{\bsnm{Howell}, \binits{J.C.}}:
\batitle{Quantum and classical coincidence imaging}.
\bjtitle{Phys. Rev. Lett.}
\bvolume{92},
\bfpage{033601}
(\byear{2004})
\doiurl{10.1103/PhysRevLett.92.033601}
\end{barticle}
\endbibitem

\bibitem[\protect\citeauthoryear{Pittman et~al.}{1995}]{Pittman95}
\begin{barticle}
\bauthor{\bsnm{Pittman}, \binits{T.B.}},
\bauthor{\bsnm{Shih}, \binits{Y.H.}},
\bauthor{\bsnm{Strekalov}, \binits{D.V.}},
\bauthor{\bsnm{Sergienko}, \binits{A.V.}}:
\batitle{Optical imaging by means of two-photon quantum entanglement}.
\bjtitle{Phys. Rev. A}
\bvolume{52},
\bfpage{3429}--\blpage{3432}
(\byear{1995})
\doiurl{10.1103/PhysRevA.52.R3429}
\end{barticle}
\endbibitem

\bibitem[\protect\citeauthoryear{Aspden et~al.}{2013}]{Aspden13}
\begin{barticle}
\bauthor{\bsnm{Aspden}, \binits{R.S.}},
\bauthor{\bsnm{Tasca}, \binits{D.S.}},
\bauthor{\bsnm{Boyd}, \binits{R.W.}},
\bauthor{\bsnm{Padgett}, \binits{M.J.}}:
\batitle{Epr-based ghost imaging using a single-photon-sensitive camera}.
\bjtitle{New Journal of Physics}
\bvolume{15}(\bissue{7}),
\bfpage{073032}
(\byear{2013})
\doiurl{10.1088/1367-2630/15/7/073032}
\end{barticle}
\endbibitem

\bibitem[\protect\citeauthoryear{Aspden et~al.}{2015}]{Aspden15}
\begin{barticle}
\bauthor{\bsnm{Aspden}, \binits{R.S.}},
\bauthor{\bsnm{Gemmell}, \binits{N.R.}},
\bauthor{\bsnm{Morris}, \binits{P.A.}},
\bauthor{\bsnm{Tasca}, \binits{D.S.}},
\bauthor{\bsnm{Mertens}, \binits{L.}},
\bauthor{\bsnm{Tanner}, \binits{M.G.}},
\bauthor{\bsnm{Kirkwood}, \binits{R.A.}},
\bauthor{\bsnm{Ruggeri}, \binits{A.}},
\bauthor{\bsnm{Tosi}, \binits{A.}},
\bauthor{\bsnm{Boyd}, \binits{R.W.}},
\bauthor{\bsnm{Buller}, \binits{G.S.}},
\bauthor{\bsnm{Hadfield}, \binits{R.H.}},
\bauthor{\bsnm{Padgett}, \binits{M.J.}}:
\batitle{Photon-sparse microscopy: visible light imaging using infrared illumination}.
\bjtitle{Optica}
\bvolume{2}(\bissue{12}),
\bfpage{1049}--\blpage{1052}
(\byear{2015})
\doiurl{10.1364/OPTICA.2.001049}
\end{barticle}
\endbibitem

\bibitem[\protect\citeauthoryear{Morris et~al.}{2015}]{Morris15}
\begin{botherref}
\oauthor{\bsnm{Morris}, \binits{P.A.}},
\oauthor{\bsnm{Aspden}, \binits{R.S.}},
\oauthor{\bsnm{Bell}, \binits{J.E.C.}},
\oauthor{\bsnm{Boyd}, \binits{R.W.}},
\oauthor{\bsnm{Padgett}, \binits{M.J.}}:
Imaging with a small number of photons.
Nature Communications
\textbf{6}(5913)
(2015)
\doiurl{10.1038/ncomms6913}
\end{botherref}
\endbibitem

\bibitem[\protect\citeauthoryear{Kalashnikov et~al.}{2014}]{PhysRevX.4.011049}
\begin{barticle}
\bauthor{\bsnm{Kalashnikov}, \binits{D.A.}},
\bauthor{\bsnm{Pan}, \binits{Z.}},
\bauthor{\bsnm{Kuznetsov}, \binits{A.I.}},
\bauthor{\bsnm{Krivitsky}, \binits{L.A.}}:
\batitle{Quantum spectroscopy of plasmonic nanostructures}.
\bjtitle{Phys. Rev. X}
\bvolume{4},
\bfpage{011049}
(\byear{2014})
\doiurl{10.1103/PhysRevX.4.011049}
\end{barticle}
\endbibitem

\bibitem[\protect\citeauthoryear{O'Sullivan et~al.}{2010}]{sullivan10pra}
\begin{barticle}
\bauthor{\bsnm{O'Sullivan}, \binits{M.N.}},
\bauthor{\bsnm{Chan}, \binits{K.W.C.}},
\bauthor{\bsnm{Boyd}, \binits{R.W.}}:
\batitle{Comparison of the signal-to-noise characteristics of quantum versus thermal ghost imaging}.
\bjtitle{Phys. Rev. A}
\bvolume{82},
\bfpage{053803}
(\byear{2010})
\doiurl{10.1103/PhysRevA.82.053803}
\end{barticle}
\endbibitem

\bibitem[\protect\citeauthoryear{Chierici}{2020}]{chierici20}
\begin{bchapter}
\bauthor{\bsnm{Chierici}, \binits{A.}}:
\bctitle{2nd scientific international conference on cbrne sicc series: 2020: book of abstracts}.
In: \bbtitle{2nd Scientific International Conference on CBRNe SICC Series},
pp. \bfpage{1}--\blpage{172}
(\byear{2020}).
\bcomment{TAB}
\end{bchapter}
\endbibitem

\bibitem[\protect\citeauthoryear{Arahata et~al.}{2021}]{Arahata21}
\begin{barticle}
\bauthor{\bsnm{Arahata}, \binits{M.}},
\bauthor{\bsnm{Mukai}, \binits{Y.}},
\bauthor{\bsnm{Cao}, \binits{B.}},
\bauthor{\bsnm{Tashima}, \binits{T.}},
\bauthor{\bsnm{Okamoto}, \binits{R.}},
\bauthor{\bsnm{Takeuchi}, \binits{S.}}:
\batitle{Wavelength variable generation and detection of photon pairs in visible and mid-infrared regions via spontaneous parametric downconversion}.
\bjtitle{J. Opt. Soc. Am. B}
\bvolume{38}(\bissue{6}),
\bfpage{1934}--\blpage{1941}
(\byear{2021})
\doiurl{10.1364/JOSAB.425550}
\end{barticle}
\endbibitem

\bibitem[\protect\citeauthoryear{Cai et~al.}{2024}]{Cai24}
\begin{barticle}
\bauthor{\bsnm{Cai}, \binits{Y.}},
\bauthor{\bsnm{Chen}, \binits{Y.}},
\bauthor{\bsnm{Dorfman}, \binits{K.}},
\bauthor{\bsnm{Xin}, \binits{X.}},
\bauthor{\bsnm{Wang}, \binits{X.}},
\bauthor{\bsnm{Huang}, \binits{K.}},
\bauthor{\bsnm{Wu}, \binits{E.}}:
\batitle{Mid-infrared single-photon upconversion spectroscopy enabled by nonlocal wavelength-to-time mapping}.
\bjtitle{Science Advances}
\bvolume{10}(\bissue{16}),
\bfpage{3503}
(\byear{2024})
\doiurl{10.1126/sciadv.adl3503}
{\href{https://arxiv.org/abs/https://www.science.org/doi/pdf/10.1126/sciadv.adl3503}{{https://www.science.org/doi/pdf/10.1126/sciadv.adl3503}}}
\end{barticle}
\endbibitem

\bibitem[\protect\citeauthoryear{Kalachev et~al.}{2008}]{Kalachev_2008}
\begin{barticle}
\bauthor{\bsnm{Kalachev}, \binits{A.A.}},
\bauthor{\bsnm{Kalashnikov}, \binits{D.A.}},
\bauthor{\bsnm{Kalinkin}, \binits{A.A.}},
\bauthor{\bsnm{Mitrofanova}, \binits{T.G.}},
\bauthor{\bsnm{Shkalikov}, \binits{A.V.}},
\bauthor{\bsnm{Samartsev}, \binits{V.V.}}:
\batitle{Biphoton spectroscopy in a strongly nondegenerate regime of spdc}.
\bjtitle{Laser Physics Letters}
\bvolume{5}(\bissue{8}),
\bfpage{600}
(\byear{2008})
\doiurl{10.1002/lapl.200810044}
\end{barticle}
\endbibitem

\bibitem[\protect\citeauthoryear{Paterova et~al.}{2020}]{Paterova20}
\begin{barticle}
\bauthor{\bsnm{Paterova}, \binits{A.V.}},
\bauthor{\bsnm{Maniam}, \binits{S.M.}},
\bauthor{\bsnm{Yang}, \binits{H.}},
\bauthor{\bsnm{Grenci}, \binits{G.}},
\bauthor{\bsnm{Krivitsky}, \binits{L.A.}}:
\batitle{Hyperspectral infrared microscopy with visible light}.
\bjtitle{Science Advances}
\bvolume{6}(\bissue{44}),
\bfpage{0460}
(\byear{2020})
\doiurl{10.1126/sciadv.abd0460}
{\href{https://arxiv.org/abs/https://www.science.org/doi/pdf/10.1126/sciadv.abd0460}{{https://www.science.org/doi/pdf/10.1126/sciadv.abd0460}}}
\end{barticle}
\endbibitem

\bibitem[\protect\citeauthoryear{Neves et~al.}{2024}]{Neves24}
\begin{botherref}
\oauthor{\bsnm{Neves}, \binits{S.}},
\oauthor{\bsnm{Kartiyasa}, \binits{A.}},
\oauthor{\bsnm{Ghosh}, \binits{S.}},
\oauthor{\bsnm{Gaulier}, \binits{G.}},
\oauthor{\bsnm{La~Volpe}, \binits{L.}},
\oauthor{\bsnm{Wolf}, \binits{J.-P.}}:
Open-path detection of organic vapors via quantum infrared spectroscopy
(2024)
\doiurl{10.48550/arXiv.2405.12822}
\end{botherref}
\endbibitem

\bibitem[\protect\citeauthoryear{Tashima et~al.}{2024}]{Tashima:24}
\begin{barticle}
\bauthor{\bsnm{Tashima}, \binits{T.}},
\bauthor{\bsnm{Mukai}, \binits{Y.}},
\bauthor{\bsnm{Arahata}, \binits{M.}},
\bauthor{\bsnm{Oda}, \binits{N.}},
\bauthor{\bsnm{Hisamitsu}, \binits{M.}},
\bauthor{\bsnm{Tokuda}, \binits{K.}},
\bauthor{\bsnm{Okamoto}, \binits{R.}},
\bauthor{\bsnm{Takeuchi}, \binits{S.}}:
\batitle{Ultra-broadband quantum infrared spectroscopy}.
\bjtitle{Optica}
\bvolume{11}(\bissue{1}),
\bfpage{81}--\blpage{87}
(\byear{2024})
\doiurl{10.1364/OPTICA.504450}
\end{barticle}
\endbibitem

\bibitem[\protect\citeauthoryear{Cardoso et~al.}{2024}]{Cardoso:24}
\begin{barticle}
\bauthor{\bsnm{Cardoso}, \binits{A.C.}},
\bauthor{\bsnm{Dong}, \binits{J.}},
\bauthor{\bsnm{Zhou}, \binits{H.}},
\bauthor{\bsnm{Joshi}, \binits{S.K.}},
\bauthor{\bsnm{Rarity}, \binits{J.G.}}:
\batitle{Methane sensing in the mid-ir using short wave ir photon counting detectors via non-linear interferometry}.
\bjtitle{Opt. Continuum}
\bvolume{3}(\bissue{5}),
\bfpage{823}--\blpage{832}
(\byear{2024})
\doiurl{10.1364/OPTCON.524280}
\end{barticle}
\endbibitem

\bibitem[\protect\citeauthoryear{Gili et~al.}{2022}]{Gili22}
\begin{barticle}
\bauthor{\bsnm{Gili}, \binits{V.F.}},
\bauthor{\bsnm{Piccinini}, \binits{C.}},
\bauthor{\bsnm{Safari~Arabi}, \binits{M.}},
\bauthor{\bsnm{Kumar}, \binits{P.}},
\bauthor{\bsnm{Besaga}, \binits{V.}},
\bauthor{\bsnm{Brambila}, \binits{E.}},
\bauthor{\bsnm{Gräfe}, \binits{M.}},
\bauthor{\bsnm{Pertsch}, \binits{T.}},
\bauthor{\bsnm{Setzpfandt}, \binits{F.}}:
\batitle{{Experimental realization of scanning quantum microscopy}}.
\bjtitle{Applied Physics Letters}
\bvolume{121}(\bissue{10}),
\bfpage{104002}
(\byear{2022})
\doiurl{10.1063/5.0095972}
{\href{https://arxiv.org/abs/https://pubs.aip.org/aip/apl/article-pdf/doi/10.1063/5.0095972/16451379/104002\_1\_online.pdf}{{https://pubs.aip.org/aip/apl/article-pdf/doi/10.1063/5.0095972/16451379/104002\_1\_online.pdf}}}
\end{barticle}
\endbibitem

\bibitem[\protect\citeauthoryear{Töpfer et~al.}{2022}]{Grafe22}
\begin{barticle}
\bauthor{\bsnm{Töpfer}, \binits{S.}},
\bauthor{\bsnm{Basset}, \binits{M.G.}},
\bauthor{\bsnm{Fuenzalida}, \binits{J.}},
\bauthor{\bsnm{Steinlechner}, \binits{F.}},
\bauthor{\bsnm{Torres}, \binits{J.P.}},
\bauthor{\bsnm{Gräfe}, \binits{M.}}:
\batitle{Quantum holography with undetected light}.
\bjtitle{Science Advances}
\bvolume{8}(\bissue{2}),
\bfpage{4301}
(\byear{2022})
\doiurl{10.1126/sciadv.abl4301}
{\href{https://arxiv.org/abs/https://www.science.org/doi/pdf/10.1126/sciadv.abl4301}{{https://www.science.org/doi/pdf/10.1126/sciadv.abl4301}}}
\end{barticle}
\endbibitem

\bibitem[\protect\citeauthoryear{Samantaray et~al.}{2017}]{Samantaray2017}
\begin{barticle}
\bauthor{\bsnm{Samantaray}, \binits{N.}},
\bauthor{\bsnm{Ruo-Berchera}, \binits{I.}},
\bauthor{\bsnm{Meda}, \binits{A.}},
\bauthor{\bsnm{Genovese}, \binits{M.}}:
\batitle{Realization of the first sub-shot-noise wide field microscope}.
\bjtitle{Light: Science {\&} Applications}
\bvolume{6}(\bissue{7}),
\bfpage{17005}--\blpage{17005}
(\byear{2017})
\doiurl{10.1038/lsa.2017.5}
\end{barticle}
\endbibitem

\bibitem[\protect\citeauthoryear{Malik et~al.}{2012}]{Malik12}
\begin{barticle}
\bauthor{\bsnm{Malik}, \binits{M.}},
\bauthor{\bsnm{Magaña-Loaiza}, \binits{O.S.}},
\bauthor{\bsnm{Boyd}, \binits{R.W.}}:
\batitle{{Quantum-secured imaging}}.
\bjtitle{Applied Physics Letters}
\bvolume{101}(\bissue{24}),
\bfpage{241103}
(\byear{2012})
\doiurl{10.1063/1.4770298}
{\href{https://arxiv.org/abs/https://pubs.aip.org/aip/apl/article-pdf/doi/10.1063/1.4770298/13566200/241103\_1\_online.pdf}{{https://pubs.aip.org/aip/apl/article-pdf/doi/10.1063/1.4770298/13566200/241103\_1\_online.pdf}}}
\end{barticle}
\endbibitem

\bibitem[\protect\citeauthoryear{Meda et~al.}{2017}]{Meda_2017}
\begin{barticle}
\bauthor{\bsnm{Meda}, \binits{A.}},
\bauthor{\bsnm{Losero}, \binits{E.}},
\bauthor{\bsnm{Samantaray}, \binits{N.}},
\bauthor{\bsnm{Scafirimuto}, \binits{F.}},
\bauthor{\bsnm{Pradyumna}, \binits{S.}},
\bauthor{\bsnm{Avella}, \binits{A.}},
\bauthor{\bsnm{Ruo-Berchera}, \binits{I.}},
\bauthor{\bsnm{Genovese}, \binits{M.}}:
\batitle{Photon-number correlation for quantum enhanced imaging and sensing}.
\bjtitle{Journal of Optics}
\bvolume{19}(\bissue{9}),
\bfpage{094002}
(\byear{2017})
\doiurl{10.1088/2040-8986/aa7b27}
\end{barticle}
\endbibitem

\bibitem[\protect\citeauthoryear{Chan et~al.}{2009}]{Chan09}
\begin{barticle}
\bauthor{\bsnm{Chan}, \binits{K.W.C.}},
\bauthor{\bsnm{O'Sullivan}, \binits{M.N.}},
\bauthor{\bsnm{Boyd}, \binits{R.W.}}:
\batitle{Two-color ghost imaging}.
\bjtitle{Phys. Rev. A}
\bvolume{79},
\bfpage{033808}
(\byear{2009})
\doiurl{10.1103/PhysRevA.79.033808}
\end{barticle}
\endbibitem

\bibitem[\protect\citeauthoryear{Shapiro}{2008}]{PhysRevA.78.061802}
\begin{barticle}
\bauthor{\bsnm{Shapiro}, \binits{J.H.}}:
\batitle{Computational ghost imaging}.
\bjtitle{Phys. Rev. A}
\bvolume{78},
\bfpage{061802}
(\byear{2008})
\doiurl{10.1103/PhysRevA.78.061802}
\end{barticle}
\endbibitem

\bibitem[\protect\citeauthoryear{D'Angelo et~al.}{2016}]{Dangelo16}
\begin{barticle}
\bauthor{\bsnm{D'Angelo}, \binits{M.}},
\bauthor{\bsnm{Pepe}, \binits{F.V.}},
\bauthor{\bsnm{Garuccio}, \binits{A.}},
\bauthor{\bsnm{Scarcelli}, \binits{G.}}:
\batitle{Correlation plenoptic imaging}.
\bjtitle{Phys. Rev. Lett.}
\bvolume{116},
\bfpage{223602}
(\byear{2016})
\doiurl{10.1103/PhysRevLett.116.223602}
\end{barticle}
\endbibitem

\bibitem[\protect\citeauthoryear{Pepe et~al.}{2017}]{Pepe17}
\begin{barticle}
\bauthor{\bsnm{Pepe}, \binits{F.V.}},
\bauthor{\bsnm{Di~Lena}, \binits{F.}},
\bauthor{\bsnm{Mazzilli}, \binits{A.}},
\bauthor{\bsnm{Edrei}, \binits{E.}},
\bauthor{\bsnm{Garuccio}, \binits{A.}},
\bauthor{\bsnm{Scarcelli}, \binits{G.}},
\bauthor{\bsnm{D'Angelo}, \binits{M.}}:
\batitle{Diffraction-limited plenoptic imaging with correlated light}.
\bjtitle{Phys. Rev. Lett.}
\bvolume{119},
\bfpage{243602}
(\byear{2017})
\doiurl{10.1103/PhysRevLett.119.243602}
\end{barticle}
\endbibitem

\bibitem[\protect\citeauthoryear{Walter et~al.}{2019}]{Walter19}
\begin{bchapter}
\bauthor{\bsnm{Walter}, \binits{D.}},
\bauthor{\bsnm{Pitsch}, \binits{C.}},
\bauthor{\bsnm{Paunescu}, \binits{G.}},
\bauthor{\bsnm{Lutzmann}, \binits{P.}}:
\bctitle{Quantum ghost imaging for remote sensing}.
In: \bbtitle{Quantum Communications and Quantum Imaging XVII}.
\bsertitle{Proc. SPIE},
vol. \bseriesno{11134},
pp. \bfpage{111340}--\blpage{17}
(\byear{2019}).
\doiurl{10.1117/12.2529268}
\end{bchapter}
\endbibitem

\bibitem[\protect\citeauthoryear{Pitsch et~al.}{2021}]{Pitsch:21}
\begin{barticle}
\bauthor{\bsnm{Pitsch}, \binits{C.}},
\bauthor{\bsnm{Walter}, \binits{D.}},
\bauthor{\bsnm{Grosse}, \binits{S.}},
\bauthor{\bsnm{Brockherde}, \binits{W.}},
\bauthor{\bsnm{B\"{u}rsing}, \binits{H.}},
\bauthor{\bsnm{Eichhorn}, \binits{M.}}:
\batitle{Quantum ghost imaging using asynchronous detection}.
\bjtitle{Appl. Opt.}
\bvolume{60}(\bissue{22}),
\bfpage{66}--\blpage{70}
(\byear{2021})
\doiurl{10.1364/AO.423634}
\end{barticle}
\endbibitem

\bibitem[\protect\citeauthoryear{Pitsch et~al.}{2023}]{Pitsch23}
\begin{barticle}
\bauthor{\bsnm{Pitsch}, \binits{C.}},
\bauthor{\bsnm{Walter}, \binits{D.}},
\bauthor{\bsnm{Gasparini}, \binits{L.}},
\bauthor{\bsnm{Bursing}, \binits{H.}},
\bauthor{\bsnm{Eichhorn}, \binits{M.}}:
\batitle{3d quantum ghost imaging}.
\bjtitle{Applied Optics}
\bvolume{62}(\bissue{23}),
\bfpage{6275}--\blpage{6281}
(\byear{2023})
\doiurl{10.1364/AO.492208}
\end{barticle}
\endbibitem

\bibitem[\protect\citeauthoryear{Scarcelli et~al.}{2003}]{Scarcelli}
\begin{barticle}
\bauthor{\bsnm{Scarcelli}, \binits{G.}},
\bauthor{\bsnm{Valencia}, \binits{A.}},
\bauthor{\bsnm{Gompers}, \binits{S.}},
\bauthor{\bsnm{Shih}, \binits{Y.}}:
\batitle{{Remote spectral measurement using entangled photons}}.
\bjtitle{Applied Physics Letters}
\bvolume{83}(\bissue{26}),
\bfpage{5560}--\blpage{5562}
(\byear{2003})
\doiurl{10.1063/1.1637131}
{\href{https://arxiv.org/abs/https://pubs.aip.org/aip/apl/article-pdf/83/26/5560/10200864/5560\_1\_online.pdf}{{https://pubs.aip.org/aip/apl/article-pdf/83/26/5560/10200864/5560\_1\_online.pdf}}}
\end{barticle}
\endbibitem

\bibitem[\protect\citeauthoryear{Yabushita and Kobayashi}{2004}]{Yabushita04}
\begin{barticle}
\bauthor{\bsnm{Yabushita}, \binits{A.}},
\bauthor{\bsnm{Kobayashi}, \binits{T.}}:
\batitle{Spectroscopy by frequency-entangled photon pairs}.
\bjtitle{Phys. Rev. A}
\bvolume{69},
\bfpage{013806}
(\byear{2004})
\doiurl{10.1103/PhysRevA.69.013806}
\end{barticle}
\endbibitem

\bibitem[\protect\citeauthoryear{Chiuri et~al.}{2022}]{PhysRevA.105.013506}
\begin{barticle}
\bauthor{\bsnm{Chiuri}, \binits{A.}},
\bauthor{\bsnm{Gianani}, \binits{I.}},
\bauthor{\bsnm{Cimini}, \binits{V.}},
\bauthor{\bsnm{De~Dominicis}, \binits{L.}},
\bauthor{\bsnm{Genoni}, \binits{M.G.}},
\bauthor{\bsnm{Barbieri}, \binits{M.}}:
\batitle{Ghost imaging as loss estimation: Quantum versus classical schemes}.
\bjtitle{Phys. Rev. A}
\bvolume{105},
\bfpage{013506}
(\byear{2022})
\doiurl{10.1103/PhysRevA.105.013506}
\end{barticle}
\endbibitem

\bibitem[\protect\citeauthoryear{Chiuri et~al.}{2023}]{doi:10.1021/acsphotonics.3c01108}
\begin{barticle}
\bauthor{\bsnm{Chiuri}, \binits{A.}},
\bauthor{\bsnm{Angelini}, \binits{F.}},
\bauthor{\bsnm{Santoro}, \binits{S.}},
\bauthor{\bsnm{Barbieri}, \binits{M.}},
\bauthor{\bsnm{Gianani}, \binits{I.}}:
\batitle{Quantum ghost imaging spectrometer}.
\bjtitle{ACS Photonics}
\bvolume{10}(\bissue{12}),
\bfpage{4299}--\blpage{4304}
(\byear{2023})
\doiurl{10.1021/acsphotonics.3c01108}
{\href{https://arxiv.org/abs/https://doi.org/10.1021/acsphotonics.3c01108}{{https://doi.org/10.1021/acsphotonics.3c01108}}}
\end{barticle}
\endbibitem

\bibitem[\protect\citeauthoryear{Leach et~al.}{2010}]{Leach10}
\begin{bchapter}
\bauthor{\bsnm{Leach}, \binits{J.}},
\bauthor{\bsnm{Jack}, \binits{B.}},
\bauthor{\bsnm{Romero}, \binits{J.}},
\bauthor{\bsnm{Ireland}, \binits{D.}},
\bauthor{\bsnm{Franke-Arnold}, \binits{S.}},
\bauthor{\bsnm{Barnett}, \binits{S.}},
\bauthor{\bsnm{Padgett}, \binits{M.}}:
\bctitle{{Quantum imaging and orbital angular momentum}}.
In: \beditor{\bsnm{Galvez}, \binits{E.J.}},
\beditor{\bsnm{Andrews}, \binits{D.L.}},
\beditor{\bsnm{Gl{\"u}ckstad}, \binits{J.}} (eds.)
\bbtitle{Proc. SPIE},
vol. \bseriesno{7613},
p. \bfpage{76130}
(\byear{2010}).
\doiurl{10.1117/12.845250}
\end{bchapter}
\endbibitem

\bibitem[\protect\citeauthoryear{Chirkin et~al.}{2018}]{Chirkin18}
\begin{barticle}
\bauthor{\bsnm{Chirkin}, \binits{A.S.}},
\bauthor{\bsnm{Gostev}, \binits{P.P.}},
\bauthor{\bsnm{Agapov}, \binits{D.P.}},
\bauthor{\bsnm{Magnitskiy}, \binits{S.A.}}:
\batitle{Ghost polarimetry: ghost imaging of polarization-sensitive objects}.
\bjtitle{Laser Physics Letters}
\bvolume{15}(\bissue{11}),
\bfpage{115404}
(\byear{2018})
\doiurl{10.1088/1612-202X/aae4a6}
\end{barticle}
\endbibitem

\bibitem[\protect\citeauthoryear{Magnitskiy et~al.}{2020}]{Magnitskiy20}
\begin{barticle}
\bauthor{\bsnm{Magnitskiy}, \binits{S.}},
\bauthor{\bsnm{Agapov}, \binits{D.}},
\bauthor{\bsnm{Chirkin}, \binits{A.}}:
\batitle{Ghost polarimetry with unpolarized pseudo-thermal light}.
\bjtitle{Opt. Lett.}
\bvolume{45}(\bissue{13}),
\bfpage{3641}--\blpage{3644}
(\byear{2020})
\doiurl{10.1364/OL.387234}
\end{barticle}
\endbibitem

\bibitem[\protect\citeauthoryear{Dong et~al.}{2016}]{Dong2016}
\begin{barticle}
\bauthor{\bsnm{Dong}, \binits{S.}},
\bauthor{\bsnm{Zhang}, \binits{W.}},
\bauthor{\bsnm{Huang}, \binits{Y.}},
\bauthor{\bsnm{Peng}, \binits{J.}}:
\batitle{Long-distance temporal quantum ghost imaging over optical fibers}.
\bjtitle{Scientific Reports}
\bvolume{6}(\bissue{1}),
\bfpage{26022}
(\byear{2016})
\doiurl{10.1038/srep26022}
\end{barticle}
\endbibitem

\bibitem[\protect\citeauthoryear{Khakimov et~al.}{2016}]{Khakimov2016}
\begin{barticle}
\bauthor{\bsnm{Khakimov}, \binits{R.I.}},
\bauthor{\bsnm{Henson}, \binits{B.M.}},
\bauthor{\bsnm{Shin}, \binits{D.K.}},
\bauthor{\bsnm{Hodgman}, \binits{S.S.}},
\bauthor{\bsnm{Dall}, \binits{R.G.}},
\bauthor{\bsnm{Baldwin}, \binits{K.G.H.}},
\bauthor{\bsnm{Truscott}, \binits{A.G.}}:
\batitle{Ghost imaging with atoms}.
\bjtitle{Nature}
\bvolume{540}(\bissue{7631}),
\bfpage{100}--\blpage{103}
(\byear{2016})
\doiurl{10.1038/nature20154}
\end{barticle}
\endbibitem

\bibitem[\protect\citeauthoryear{Chan et~al.}{2009}]{Chan:09}
\begin{barticle}
\bauthor{\bsnm{Chan}, \binits{K.W.C.}},
\bauthor{\bsnm{O'Sullivan}, \binits{M.N.}},
\bauthor{\bsnm{Boyd}, \binits{R.W.}}:
\batitle{High-order thermal ghost imaging}.
\bjtitle{Opt. Lett.}
\bvolume{34}(\bissue{21}),
\bfpage{3343}--\blpage{3345}
(\byear{2009})
\doiurl{10.1364/OL.34.003343}
\end{barticle}
\endbibitem

\bibitem[\protect\citeauthoryear{Chiuri et~al.}{2024}]{PhysRevA.109.042617}
\begin{barticle}
\bauthor{\bsnm{Chiuri}, \binits{A.}},
\bauthor{\bsnm{Barbieri}, \binits{M.}},
\bauthor{\bsnm{Venditti}, \binits{I.}},
\bauthor{\bsnm{Angelini}, \binits{F.}},
\bauthor{\bsnm{Battocchio}, \binits{C.}},
\bauthor{\bsnm{Paris}, \binits{M.G.A.}},
\bauthor{\bsnm{Gianani}, \binits{I.}}:
\batitle{Fast remote spectral discrimination through ghost spectrometry}.
\bjtitle{Phys. Rev. A}
\bvolume{109},
\bfpage{042617}
(\byear{2024})
\doiurl{10.1103/PhysRevA.109.042617}
\end{barticle}
\endbibitem

\bibitem[\protect\citeauthoryear{Chowdhury et~al.}{2021}]{Chowdhury21}
\begin{bchapter}
\bauthor{\bsnm{Chowdhury}, \binits{A.R.}},
\bauthor{\bsnm{Burney}, \binits{U.}},
\bauthor{\bsnm{Hutter}, \binits{D.}},
\bauthor{\bsnm{Lee}, \binits{T.-A.}},
\bauthor{\bsnm{Hutter}, \binits{T.}}:
\bctitle{Detection of toxic chemicals in hand sanitizers using near-infrared spectroscopy}.
In: \bbtitle{OSA Optical Sensors and Sensing Congress 2021 (AIS, FTS, HISE, SENSORS, ES)},
pp. \bfpage{6}--\blpage{5}
(\byear{2021}).
\doiurl{10.1364/AIS.2021.AW6E.5}
\end{bchapter}
\endbibitem

\bibitem[\protect\citeauthoryear{Czechlowski et~al.}{2019}]{CZECHLOWSKI201914}
\begin{barticle}
\bauthor{\bsnm{Czechlowski}, \binits{M.}},
\bauthor{\bsnm{Marcinkowski}, \binits{D.}},
\bauthor{\bsnm{Golimowska}, \binits{R.}},
\bauthor{\bsnm{Berger}, \binits{W.A.}},
\bauthor{\bsnm{Golimowski}, \binits{W.}}:
\batitle{Spectroscopy approach to methanol detection in waste fat methyl esters}.
\bjtitle{Spectrochimica Acta Part A: Molecular and Biomolecular Spectroscopy}
\bvolume{210},
\bfpage{14}--\blpage{20}
(\byear{2019})
\doiurl{10.1016/j.saa.2018.11.003}
\end{barticle}
\endbibitem

\bibitem[\protect\citeauthoryear{Ionescu and Doctorala}{2018}]{DCM}
\begin{bchapter}
\bauthor{\bsnm{Ionescu}, \binits{M.}},
\bauthor{\bsnm{Doctorala}, \binits{S.}}:
\bctitle{Measuring and detecting blood glucose by methods non-invasive}.
In: \bbtitle{2018 10th International Conference on Electronics, Computers and Artificial Intelligence (ECAI)},
pp. \bfpage{1}--\blpage{7}
(\byear{2018}).
\doiurl{10.1109/ECAI.2018.8678943}
\end{bchapter}
\endbibitem

\bibitem[\protect\citeauthoryear{Swann and Gilbert}{2000}]{Swann00}
\begin{barticle}
\bauthor{\bsnm{Swann}, \binits{W.C.}},
\bauthor{\bsnm{Gilbert}, \binits{S.L.}}:
\batitle{Pressure-induced shift and broadening of 1510--1540-nm acetylene wavelength calibration lines}.
\bjtitle{J. Opt. Soc. Am. B}
\bvolume{17}(\bissue{7}),
\bfpage{1263}--\blpage{1270}
(\byear{2000})
\doiurl{10.1364/JOSAB.17.001263}
\end{barticle}
\endbibitem

\bibitem[\protect\citeauthoryear{Gilbert and Swann}{2001}]{nist2001}
\begin{botherref}
\oauthor{\bsnm{Gilbert}, \binits{S.L.}},
\oauthor{\bsnm{Swann}, \binits{W.C.}}:
Standard reference materials : Acetylene 12c2h2 absorption reference for 1510 nm to 1540 nm wavelength calibration - srm 2517a,
pp. 1--5.
National Institute of Standards and Technology,
Gaithersburg, MD
(2001)
\end{botherref}
\endbibitem

\bibitem[\protect\citeauthoryear{}{}]{hitran}
\begin{botherref}
Hitran database.
\url{https://hitran.iao.ru/home}
\end{botherref}
\endbibitem

\bibitem[\protect\citeauthoryear{Shen}{1981}]{Shen81}
\begin{barticle}
\bauthor{\bsnm{Shen}, \binits{T.T.}}:
\batitle{Control techniques for gas emissions from hazardous waste landfills}.
\bjtitle{Journal of the Air Pollution Control Association}
\bvolume{31}(\bissue{2}),
\bfpage{132}--\blpage{135}
(\byear{1981})
\doiurl{10.1080/00022470.1981.10465200}
{\href{https://arxiv.org/abs/https://doi.org/10.1080/00022470.1981.10465200}{{https://doi.org/10.1080/00022470.1981.10465200}}}
\end{barticle}
\endbibitem

\bibitem[\protect\citeauthoryear{Fleisher et~al.}{2017}]{Fleisher2017}
\begin{barticle}
\bauthor{\bsnm{Fleisher}, \binits{A.J.}},
\bauthor{\bsnm{Long}, \binits{D.A.}},
\bauthor{\bsnm{Liu}, \binits{Q.}},
\bauthor{\bsnm{Gameson}, \binits{L.}},
\bauthor{\bsnm{Hodges}, \binits{J.T.}}:
\batitle{Optical measurement of radiocarbon below unity fraction modern by linear absorption spectroscopy}.
\bjtitle{The Journal of Physical Chemistry Letters}
\bvolume{8}(\bissue{18}),
\bfpage{4550}--\blpage{4556}
(\byear{2017})
\doiurl{10.1021/acs.jpclett.7b02105}
\end{barticle}
\endbibitem

\bibitem[\protect\citeauthoryear{Romaniello et~al.}{2021}]{rs13224502}
\begin{botherref}
\oauthor{\bsnm{Romaniello}, \binits{V.}},
\oauthor{\bsnm{Spinetti}, \binits{C.}},
\oauthor{\bsnm{Silvestri}, \binits{M.}},
\oauthor{\bsnm{Buongiorno}, \binits{M.F.}}:
A methodology for co2 retrieval applied to hyperspectral prisma data.
Remote Sensing
\textbf{13}(22)
(2021)
\doiurl{10.3390/rs13224502}
\end{botherref}
\endbibitem

\bibitem[\protect\citeauthoryear{Li~Yoon et~al.}{1999}]{A903398J}
\begin{barticle}
\bauthor{\bsnm{Li~Yoon}, \binits{W.}},
\bauthor{\bsnm{D.~Jee}, \binits{R.}},
\bauthor{\bsnm{C.~Moffat}, \binits{A.}},
\bauthor{\bsnm{D.~Blackler}, \binits{P.}},
\bauthor{\bsnm{Yeung}, \binits{K.}},
\bauthor{\bsnm{C.~Lee}, \binits{D.}}:
\batitle{Construction and transferability of a spectral library for the identification of common solvents by near-infrared transflectance spectroscopy}.
\bjtitle{Analyst}
\bvolume{124},
\bfpage{1197}--\blpage{1203}
(\byear{1999})
\doiurl{10.1039/A903398J}
\end{barticle}
\endbibitem

\bibitem[\protect\citeauthoryear{Bruschini et~al.}{2019}]{Bruschini2019}
\begin{barticle}
\bauthor{\bsnm{Bruschini}, \binits{C.}},
\bauthor{\bsnm{Homulle}, \binits{H.}},
\bauthor{\bsnm{Antolovic}, \binits{I.M.}},
\bauthor{\bsnm{Burri}, \binits{S.}},
\bauthor{\bsnm{Charbon}, \binits{E.}}:
\batitle{Single-photon avalanche diode imagers in biophotonics: review and outlook}.
\bjtitle{Light: Science {\&} Applications}
\bvolume{8}(\bissue{1}),
\bfpage{87}
(\byear{2019})
\doiurl{10.1038/s41377-019-0191-5}
\end{barticle}
\endbibitem

\bibitem[\protect\citeauthoryear{Madonini et~al.}{2021}]{Madonini21}
\begin{barticle}
\bauthor{\bsnm{Madonini}, \binits{F.}},
\bauthor{\bsnm{Severini}, \binits{F.}},
\bauthor{\bsnm{Zappa}, \binits{F.}},
\bauthor{\bsnm{Villa}, \binits{F.}}:
\batitle{Single photon avalanche diode arrays for quantum imaging and microscopy}.
\bjtitle{Advanced Quantum Technologies}
\bvolume{4}(\bissue{7}),
\bfpage{2100005}
(\byear{2021})
\doiurl{10.1002/qute.202100005}
{\href{https://arxiv.org/abs/https://onlinelibrary.wiley.com/doi/pdf/10.1002/qute.202100005}{{https://onlinelibrary.wiley.com/doi/pdf/10.1002/qute.202100005}}}
\end{barticle}
\endbibitem

\bibitem[\protect\citeauthoryear{Fisher-Levine and Nomerotski}{2016}]{Fisher-Levine_2016}
\begin{barticle}
\bauthor{\bsnm{Fisher-Levine}, \binits{M.}},
\bauthor{\bsnm{Nomerotski}, \binits{A.}}:
\batitle{Timepixcam: a fast optical imager with time-stamping}.
\bjtitle{Journal of Instrumentation}
\bvolume{11}(\bissue{03}),
\bfpage{03016}
(\byear{2016})
\doiurl{10.1088/1748-0221/11/03/C03016}
\end{barticle}
\endbibitem

\bibitem[\protect\citeauthoryear{Nomerotski}{2019}]{NOMEROTSKI201926}
\begin{barticle}
\bauthor{\bsnm{Nomerotski}, \binits{A.}}:
\batitle{Imaging and time stamping of photons with nanosecond resolution in timepix based optical cameras}.
\bjtitle{Nuclear Instruments and Methods in Physics Research Section A: Accelerators, Spectrometers, Detectors and Associated Equipment}
\bvolume{937},
\bfpage{26}--\blpage{30}
(\byear{2019})
\doiurl{10.1016/j.nima.2019.05.034}
\end{barticle}
\endbibitem

\bibitem[\protect\citeauthoryear{Ianzano et~al.}{2020}]{Ianzano2020}
\begin{barticle}
\bauthor{\bsnm{Ianzano}, \binits{C.}},
\bauthor{\bsnm{Svihra}, \binits{P.}},
\bauthor{\bsnm{Flament}, \binits{M.}},
\bauthor{\bsnm{Hardy}, \binits{A.}},
\bauthor{\bsnm{Cui}, \binits{G.}},
\bauthor{\bsnm{Nomerotski}, \binits{A.}},
\bauthor{\bsnm{Figueroa}, \binits{E.}}:
\batitle{Fast camera spatial characterization of photonic polarization entanglement}.
\bjtitle{Scientific Reports}
\bvolume{10}(\bissue{1}),
\bfpage{6181}
(\byear{2020})
\doiurl{10.1038/s41598-020-62020-z}
\end{barticle}
\endbibitem

\bibitem[\protect\citeauthoryear{Vidyapin et~al.}{2023}]{Vidyapin2023}
\begin{barticle}
\bauthor{\bsnm{Vidyapin}, \binits{V.}},
\bauthor{\bsnm{Zhang}, \binits{Y.}},
\bauthor{\bsnm{England}, \binits{D.}},
\bauthor{\bsnm{Sussman}, \binits{B.}}:
\batitle{Characterisation of a single photon event camera for quantum imaging}.
\bjtitle{Scientific Reports}
\bvolume{13}(\bissue{1}),
\bfpage{1009}
(\byear{2023})
\doiurl{10.1038/s41598-023-27842-7}
\end{barticle}
\endbibitem

\bibitem[\protect\citeauthoryear{Resta et~al.}{2023}]{Resta2023}
\begin{barticle}
\bauthor{\bsnm{Resta}, \binits{G.V.}},
\bauthor{\bsnm{Stasi}, \binits{L.}},
\bauthor{\bsnm{Perrenoud}, \binits{M.}},
\bauthor{\bsnm{El-Khoury}, \binits{S.}},
\bauthor{\bsnm{Brydges}, \binits{T.}},
\bauthor{\bsnm{Thew}, \binits{R.}},
\bauthor{\bsnm{Zbinden}, \binits{H.}},
\bauthor{\bsnm{Bussi{\`e}res}, \binits{F.}}:
\batitle{Gigahertz detection rates and dynamic photon-number resolution with superconducting nanowire arrays}.
\bjtitle{Nano Letters}
\bvolume{23}(\bissue{13}),
\bfpage{6018}--\blpage{6026}
(\byear{2023})
\doiurl{10.1021/acs.nanolett.3c01228}
\end{barticle}
\endbibitem

\bibitem[\protect\citeauthoryear{Hao et~al.}{2024}]{Hao2024}
\begin{barticle}
\bauthor{\bsnm{Hao}, \binits{H.}},
\bauthor{\bsnm{Zhao}, \binits{Q.-Y.}},
\bauthor{\bsnm{Huang}, \binits{Y.-H.}},
\bauthor{\bsnm{Deng}, \binits{J.}},
\bauthor{\bsnm{Yang}, \binits{F.}},
\bauthor{\bsnm{Ru}, \binits{S.-Y.}},
\bauthor{\bsnm{Liu}, \binits{Z.}},
\bauthor{\bsnm{Wan}, \binits{C.}},
\bauthor{\bsnm{Liu}, \binits{H.}},
\bauthor{\bsnm{Li}, \binits{Z.-J.}},
\bauthor{\bsnm{Wang}, \binits{H.-B.}},
\bauthor{\bsnm{Tu}, \binits{X.-C.}},
\bauthor{\bsnm{Zhang}, \binits{L.-B.}},
\bauthor{\bsnm{Jia}, \binits{X.-Q.}},
\bauthor{\bsnm{Wu}, \binits{X.-L.}},
\bauthor{\bsnm{Chen}, \binits{J.}},
\bauthor{\bsnm{Kang}, \binits{L.}},
\bauthor{\bsnm{Wu}, \binits{P.-H.}}:
\batitle{A compact multi-pixel superconducting nanowire single-photon detector array supporting gigabit space-to-ground communications}.
\bjtitle{Light: Science {\&} Applications}
\bvolume{13}(\bissue{1}),
\bfpage{25}
(\byear{2024})
\doiurl{10.1038/s41377-023-01374-1}
\end{barticle}
\endbibitem

\bibitem[\protect\citeauthoryear{Wollman et~al.}{2019}]{Wollman:19}
\begin{barticle}
\bauthor{\bsnm{Wollman}, \binits{E.E.}},
\bauthor{\bsnm{Verma}, \binits{V.B.}},
\bauthor{\bsnm{Lita}, \binits{A.E.}},
\bauthor{\bsnm{Farr}, \binits{W.H.}},
\bauthor{\bsnm{Shaw}, \binits{M.D.}},
\bauthor{\bsnm{Mirin}, \binits{R.P.}},
\bauthor{\bsnm{Nam}, \binits{S.W.}}:
\batitle{Kilopixel array of superconducting nanowire single-photon detectors}.
\bjtitle{Opt. Express}
\bvolume{27}(\bissue{24}),
\bfpage{35279}--\blpage{35289}
(\byear{2019})
\doiurl{10.1364/OE.27.035279}
\end{barticle}
\endbibitem

\end{thebibliography}

\end{document}